\documentclass[journal,10pt]{IEEEtran}

\usepackage{cite}
\usepackage{color}
\ifCLASSINFOpdf
   \usepackage[pdftex]{graphicx}
\else
  \usepackage[dvips]{graphicx}
   \DeclareGraphicsExtensions{.eps}
\fi
\usepackage{subfigure}

\usepackage{amsmath}
\usepackage{amsfonts,helvet}
\usepackage{amssymb}

\usepackage{algorithm}
\usepackage{algpseudocode}
\usepackage{array}
\usepackage{multirow}

\begin{document}

%\title{Real-Time Implementation and Characterization for Compact Full Duplex MIMO Radios}
\title{Compact Full Duplex MIMO Radios in\\ D2D Underlaid Cellular Networks:\\From System Design to Prototype Results}

\author{MinKeun Chung,~\IEEEmembership{Member,~IEEE}, Min Soo Sim,~\IEEEmembership{Student Member,~IEEE},\\ Dong Ku Kim,~\IEEEmembership{Senior Member,~IEEE}, and Chan-Byoung Chae,~\IEEEmembership{Senior Member,~IEEE}

\thanks{M. Chung and M. S. Sim, and C.-B. Chae (corresponding author) are with the School of Integrated Technology, Yonsei University, Korea (E-mail: \{minkeun.chung, simms, cbchae\}@yonsei.ac.kr). D. K. Kim is with the School of Electrical and Electronic Engineering, Yonsei University, Korea (E-mail: dkkim@yonsei.ac.kr).}
%\thanks{Y. Fang is with the Department of Electrical and Computer Engineering, University of Florida, Gainesville, FL, 32611, USA e-mail: (see http://winet.ece.ufl.edu/tvt/ for further information regarding IEEE TVT.)}% <-this % stops a space
%\thanks{Manuscript received XXX, XX, 2015; revised XXX, XX, 2015.}
}

%\markboth{IEEE Transactions on Vehicular Technology,~Vol.~XX, No.~XX, XXX~2016}
{}
%{Shell \MakeLowercase{\textit{et al.}}: Bare Demo of IEEEtran.cls for Journals}

\maketitle

%%%%%%%%%%%%%%%%%%%%%%%%%%%%%%%%%
%%%%%%%%%%%%%%%%%%%%%%%%%%%%%%%%%
\begin{abstract}
This paper considers the implementation and application possibilities of a compact full duplex multiple-input multiple-output~(MIMO) architecture where direct communication exists between users, e.g., device-to-device~(D2D) and cellular link coexisting on the same spectrum. For the architecture of the compact full duplex radio, we combine an analog self-interference canceler based dual-polarization with high cross-polarization discrimination (XPD) and Long Term Evolution (LTE)-based per-subcarrier digital self-interference canceler. While we consider the compactness and power efficiency of an analog solution, we focus on the digital canceler design with robustness to a frequency-selective channel and high compatibility with a conventional LTE system. For an over-the-air wireless experiment of full duplex testbed with a two-user-pair, we implement a full duplex MIMO physical layer (PHY), supporting 20 MHz bandwidth, on an FPGA-based software-defined radio platform. Further, we propose a novel timing synchronization method to construct a more viable full duplex MIMO link. By having the full duplex link prototype fully operating in  real-time, we present the first characterization of the proposed compact full duplex MIMO performance depending on the transmit power of the full duplex node. We also show the link quality between nodes. One of the crucial insights of this work is that the full duplex operation of a user is capable of acquiring the throughput gain if the user has self-interference capability with guaranteed performance.

\end{abstract}


\IEEEpeerreviewmaketitle

%%%%%%%%%%%%%%%%%%%%%%%%%%%%%%%%%
%%%%%%%%%%%%%%%%%%%%%%%%%%%%%%%%%
\section{Introduction}
A particular technology designed to boost throughput in wireless communications and that many researchers consider to be novel is a technology known as in-band full duplex~\cite{Sabharwal2014},~\cite{hong2014}. In theory it doubles the throughput by simultaneous transmission and reception of the same frequency in a wireless radio.  Researchers~\cite{Jchoi2010, Mjain2011, Duarte2010, Duarte2012, Duarte2014, Hong2012, Bharadia2013, Bharadia2014, Choi2013, Thuusari2015, Mchung2015, aryafar2012} have demonstrated that its theoretical potential can be transformed into practical achievement for new spectrum opportunities in terms of full duplex. Doubling throughput is of course a desirable achievement, but the challenge lies in attaining full duplex communication within the same spectrum is challenging. Specifically, the challenge specifically is managing interference. Interference does not occur in conventional half duplex communications. The key obstacle to realizing full duplex communications is the high-power differential between the transmitter's echo, the so-called self-interference, and the desired signal from its partner. As wireless signals attenuate quickly over distance, unwanted self-interference from an own transmitting antenna is billions of times (100 dB+) stronger than the desired receive signal from the partner. As a consequence of the high power differential, the desired receive signal is, without intelligent interference management strategies, flooded with self-interference. Self interference differs from general interference in that the original information of interference is known at the receiving position. Such knowledge could be considered a sufficient benefit for self-interference suppression. But a tremendous amount of self-interference requires several phases for suppression over analog and digital domains. Despite the best efforts of self-interference cancelers, the result is an unwieldy transceiver, for the cancelers operate in several phases to maximize the cancellation performance. Hence, the researchers consider the main candidate of full duplex radio is considered to be at the base station~(BS). 

\begin{figure*}[h]
    \centering
    \includegraphics[width = 7in]{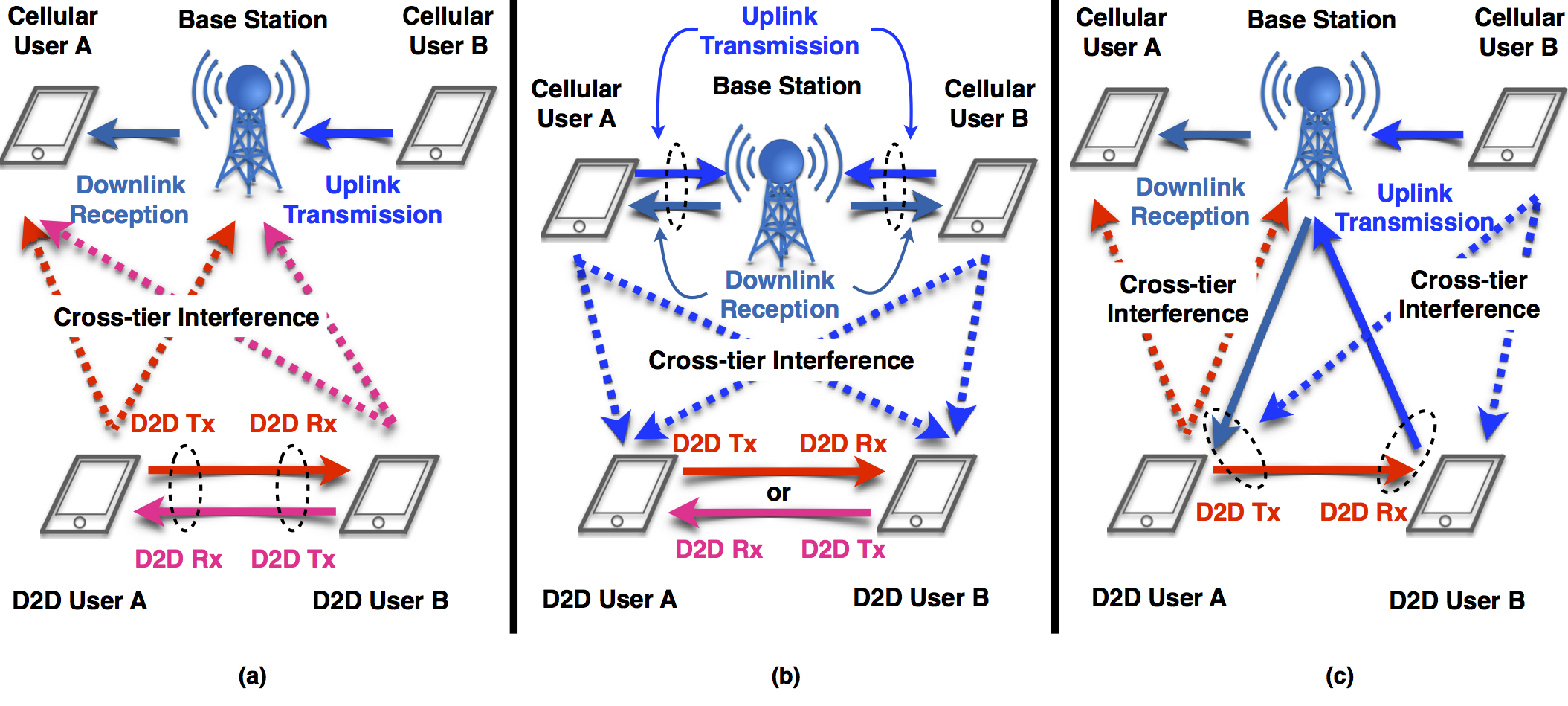}
    \caption{Full duplex scenarios in a cellular network coexisting with D2D link within the same shared cellular spectrum: (a) power-controlled full duplex D2D system (b) uplink power-controlled full duplex cellular system (c) power-controlled full duplex D2D system underlying the downlink cellular network \& full duplex D2D system underlying the uplink power-controlled cellular network. In all scenarios, we assume that a BS in a single-cell could operate on half or full duplex mode. The dashed-circles and the dotted lines in each subfigure represent power-controlled full duplex operations at the user node and cross-tier interference from the users between D2D and cellular link, respectively.}              
    \label{fig:d2dcellular}
    \end{figure*}

In recent years, researchers~\cite{Klaus2009,Corson2010, Gabor2012} have proposed direct communication between mobile users, e.g., device-to-device~(D2D) communication, as a means of improving spectral efficiency, lowering latency, and reducing handset power consumption. A network in which D2D and cellular links coexist on a common spectrum suggest various scenarios regarding the full duplex operation of a user. Conceptions of such scenarios are illustrated in three components as illustrated in Fig.~\ref{fig:d2dcellular}. Each component is explained in more detail in Section~\ref{sec:background}. 

To prevent the throughput performance-degradation to existing cellular links, researchers~\cite{Janis2009, Lei2012, Lee2015} have proposed a power control technique for D2D. As a way of exploiting power control, the authors in~\cite{Ali2014, Cheng2014, Zhang2015} investigated the applicability of full duplex D2D underlaid cellular networks. If mobile users have a self-interference cancelling capability depending on the range of power control, the merits of direct communications between mobile users would create a synergistic effect with full duplex. Therefore, full duplex communication between mobile users could be considered as an interesting candidate to boost link or system throughput. To design mobile radio to operate in full duplex mode, one should consider-- with guaranteed self-interference cancelling performance-- the following characteristics: simplicity, compactness, power- and cost-efficiencies. 

    \begin{figure*}[h]
    \centering
    \includegraphics[width = 5in]{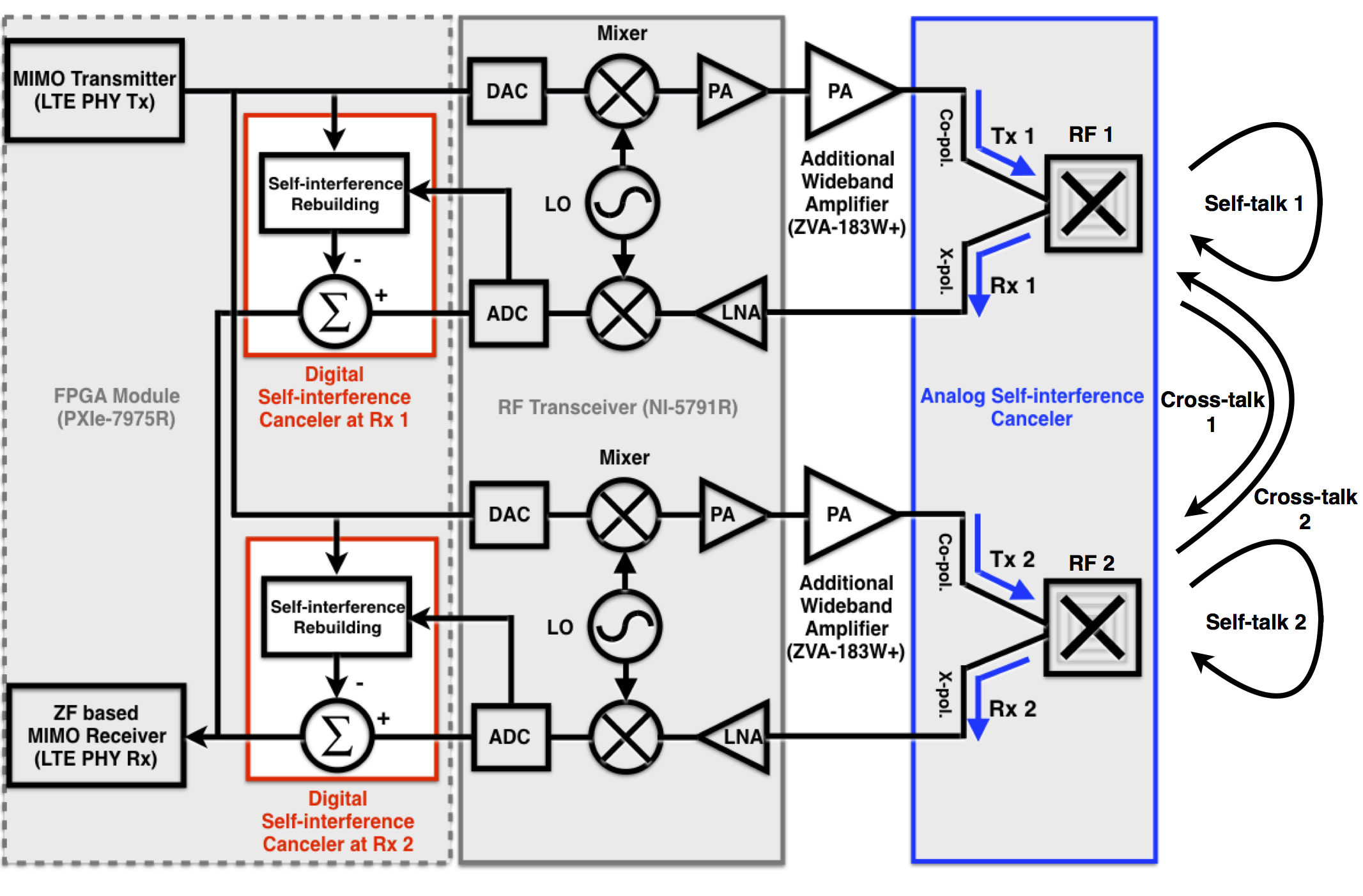}
    \caption{Block diagram of the proposed compact full duplex MIMO radio architecture. }              
    \label{fig:MIMO_FD}
    \end{figure*}

One attractive method for multiplying the capacity of a radio link is multiple-input multiple-output~(MIMO). The combination of MIMO and full duplex, however. gives rise to high processing complexity. Specifically, an $N$-dimensional full duplex radio generates a number of  $N^2$ self-interference sources. For mobile users, these sources could prove to be an intolerable burden. In this paper, we propose an architecture that combines the physical layer~(PHY) design of two powerful technologies -- full duplex and MIMO. This architecture gives consideration to compactness and power efficiency for user operation. With real-time implementation of the proposed architecture, we seek to characterize the link throughput performance with compact full duplex MIMO radios according to the transmit power of a full duplex node and the link quality.

%\hfill August 4, 2015
%%%%%%%%%%%%%%%%%%%%%%%%%%%%%%%%%
%%%%%%%%%%%%%%%%%%%%%%%%%%%%%%%%%
\subsection{Related Work}
There has been considerable interest in finding practical solutions to the challenge of realizing full duplex radio. The simplest solutions were introduced in~\cite{Jchoi2010, Mjain2011, Duarte2010} for antenna-isolation techniques; these exploit the antenna spacing and positioning between the transmit~(Tx) and receive~(Rx) antennas. Using only passive cancellation methods, however, yielded only limited results in terms of satisfying the high requirement of self-interference cancellation. The authors in~\cite{Jchoi2010, Mjain2011, Duarte2010, Duarte2012, Duarte2014} proposed techniques that employed active cancellation. The basic idea in these works was to tap the Tx signal over an additional transmission line in the analog domain for cancellation~\cite{Jchoi2010, Mjain2011}, and use an extra Tx chain to generate a modified copy of the Tx signal~\cite{Duarte2010, Duarte2012, Duarte2014}. To integrate radio frequency~(RF) front-end in the limited area of a transceiver that allows simultaneous transmission and reception, researchers~\cite{Hong2012} proposed a single shared antenna that uses a circulator, thereby avoiding the need for antenna separation or positioning between the Tx and Rx antennas. The problem with this idea, though, is that a circulator typically provides limited isolation between the Tx and Rx ports, the isolation of which is about 15-20 dB. The solutions that use the circulator for self-interference suppression acquire active cancellation through more sophisticated techniques~\cite{Bharadia2013, Bharadia2014, Choi2013, Thuusari2015}. The main idea is to design a dedicated analog cancellation circuit based on a control algorithm~\cite{Bharadia2013, Bharadia2014} and an adaptive filtering algorithm~\cite{Choi2013, Thuusari2015}. The authors in~\cite{Debaillie2014, Everett2014} introduced the dual-polarized antenna as another single RF front-end for full duplex operation. In~\cite{Mchung2015}, the researchers demonstrated a real-time full duplex single-input single-output (SISO) radio system based on a dual-polarization antenna. In an effort to combine both full duplex and MIMO, the researchers in~\cite{aryafar2012} studied implementation-based antenna spacing and positioning. The researchers in~\cite{Bharadia2014} proposed a full duplex MIMO radio based on the work done in~\cite{Bharadia2013}.

For full duplex operation at the user node, though, the aforementioned studies~\cite{Jchoi2010, Mjain2011, Duarte2010, Duarte2012, Duarte2014, Hong2012, Bharadia2013, Bharadia2014, Choi2013, Thuusari2015} share one main limitation-- none of them satisfactorily meet the needs for compactness and power efficiency. Thus the goal of the current research is twofold-- 1) to present the implementation of a compact and power-efficient full duplex MIMO prototype for a user and 2) to show, using our prototype, the application possibilities through an over-the-air wireless test.

%%%%%%%%%%%%%%%%%%%%%%%%%%%%%%%%%
%%%%%%%%%%%%%%%%%%%%%%%%%%%%%%%%%
\subsection{Contributions}
The main contribution of this paper is that it offers the first characterization of compact full duplex MIMO performances through real-time prototyping with an analog canceler, digital canceler, and demodulator for the desired signal. Prior to implementing of full duplex MIMO radios, we prototype a Long Term Evolution~(LTE)-based conventional half duplex PHY. The prototype will provide a performance benchmark to clearly compare the throughput improvement of our compact full duplex MIMO radios, on a software-defined radio~(SDR) platform. Presented as another benchmark in this work is the measurement result of the full duplex SISO prototype implemented on the same SDR platform. Our compact full duplex MIMO prototype is implemented in a~$2 \times 2$~configuration. As illustrated in Fig.~\ref{fig:MIMO_FD}, the configuration consists of two Tx ports and two Rx ports. To enable full duplex operation at the user node, we consider the dual-polarization-based RF front-end with high cross-polarization discrimination (XPD) for analog self-interference cancellation. The merit of the dual-polarization-based antenna with high XPD as a full duplex RF solution (in addition to the single-shared antenna in the circulator-based solution~\cite{Hong2012,Bharadia2013, Bharadia2014, Choi2013, Thuusari2015}) offers one main merit; it provides, through solely passive isolation, fairly good self-interference cancellation capability without needing to consume additional power consumption. As shown in Fig.~\ref{fig:MIMO_FD}, we construct a full duplex~$2 \times 2$~MIMO RF using two dual-polarized antennas. In order to suppress residual self-interference and demodulate the desired signal after passing the dual-polarized antenna even in a mobile environment, we jointly design per-subcarrier digital self-interference cancelers on an LTE-based MIMO PHY prototype that includes a demodulator. 

With this prototype, we study an over-the-air time synchronization method in full duplex systems. Using a modified LTE frame structure, we introduce a novel timing synchronization algorithm for full duplex communications. Through implementation of the proposed protocol on our full duplex MIMO prototype, we provide field test results of our proposed timing synchronization algorithm in a real-world wireless channel.

For experimental investigation through our compact full duplex link prototype with a two-user pair, we measure self-interference cancellation capability according to the transmit power of the full duplex node. Further, with the measurement campaign for link quality from several fixed nodes, we derive link throughput characteristics that depend on link quality and the transmit power of the full duplex node.

This paper is organized as follows. Section~\ref{sec:background} describes the background of compact full duplex radio. Section~\ref{sec:analogsic} provides the characteristic of dual-polarization-based full duplex MIMO front-end. Section~\ref{sec:fullmimophy} presents the design of the digital self-interference canceler. Section~\ref{sec:prototypesetup} briefly describes our prototype setup. Section~\ref{sec:phyeval} describes the measurement campaign for self-interference cancellation level and link quality while also providing the evaluation results. Section~\ref{sec:conclusion} lays out our conclusions. 

Throughout this paper, the unit used regarding analog/digital self-interference cancellation level $\alpha$ and $\delta$ is decibel (dB). In addition, time and subcarrier indices are represented by $n$ and $k$, respectively.

%%%%%%%%%%%%%%%%%%%%%%%%%%%%%%%%%
%%%%%%%%%%%%%%%%%%%%%%%%%%%%%%%%%
\section{Background}
\label{sec:background}
As background to this paper, we first describe the appropriate scenarios in which it is desirable to improve network throughput by a user having the capacity for self-interference cancellation. Then, to address these scenarios, we derive the necessary requirements built on system specifications. The requirements shown in this section are provided as a criterion for designing the proposed compact full duplex MIMO radio.

    \begin{figure*}
    \centering
    \includegraphics[width = 6in]{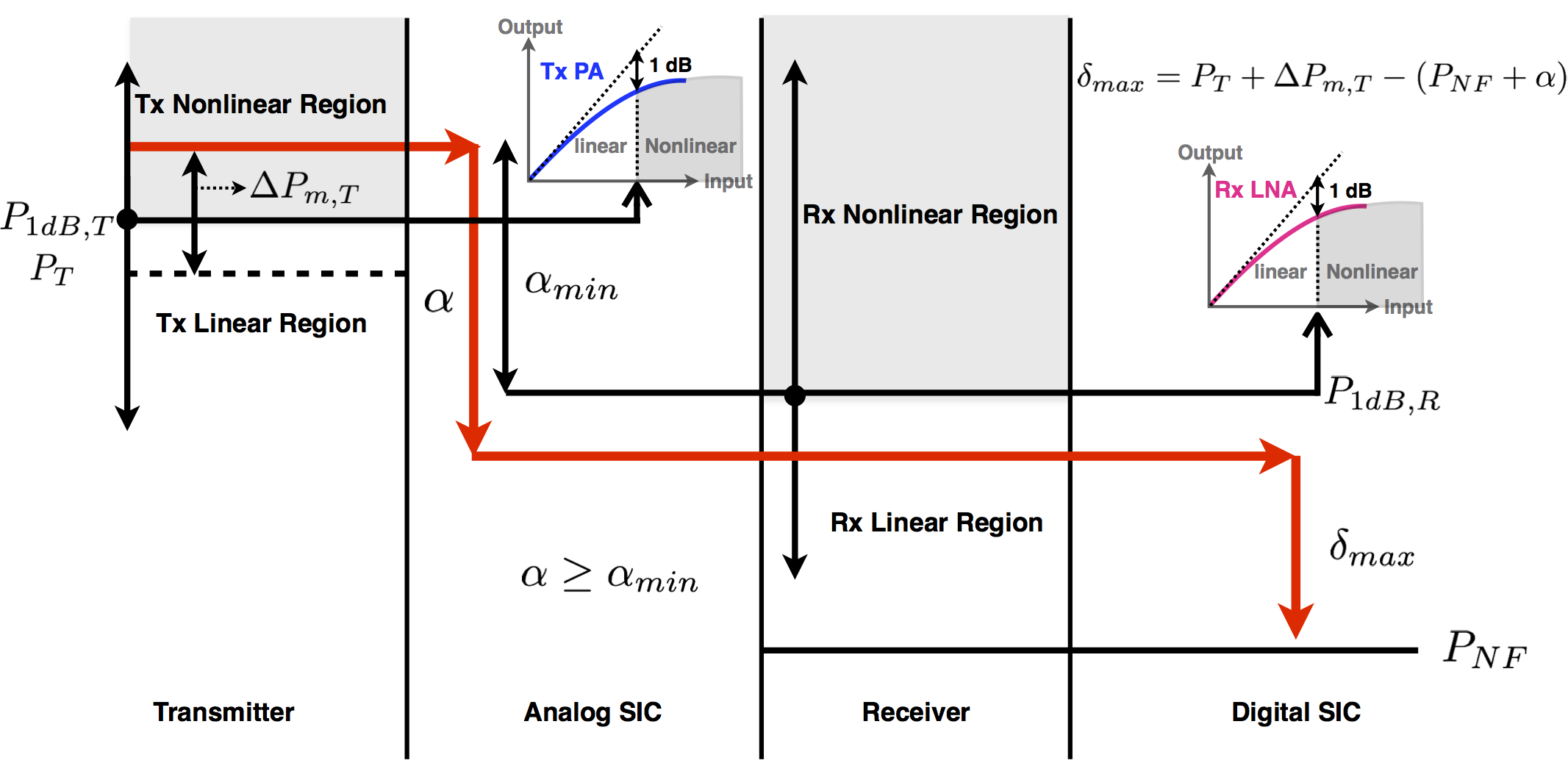}
    \caption{Requirements for self-interference cancellation under a given system specification. We consider 1-dB compression points ($P_{1 {\rm{dB}},T}$, $P_{1 {\rm{dB}},R}$) of PA and LNA in the Tx and Rx chains as key parameters for self-interference cancellation strategy with low complexity in the user node. The red arrows in figure represent the route of our strategy to minimize the impact of nonlinearity. The basic idea is that PA nonlinearity is avoided by the scheduling of a power-controlled user, LNA nonlinearity is avoided by passive analog canceler guaranteeing $\alpha_{min}$ at  least.}              
    \label{fig:Requirements}
    \end{figure*}

%%%%%%%%%%%%%%%%%%%%%%%%%%%%%%%%%
%%%%%%%%%%%%%%%%%%%%%%%%%%%%%%%%%
\subsection{Network Scenarios}
We consider a single-cell interference network in which, as illustrated in Fig.~\ref{fig:d2dcellular}, direct D2D and cellular link coexist on the common spectrum. In this network, D2D and cellular link mutually experience cross-tier interference due to the transmissions in each link. Accordingly, power control is utilized to curb the severe interference between the D2D and cellular links. If at the user node, it is possible to cancel self-interference cancellation against the controlled transmit power, then D2D or a cellular user could simultaneously transmit and receive instead of time-division duplex~(TDD). In Fig.~\ref{fig:d2dcellular}, full duplex operation is signified with the dashed-circles and presented with several network scenarios as follows.

\subsubsection{Power-controlled full duplex D2D system}
Consider a scenario, as depicted in Fig.~\ref{fig:d2dcellular}-(a), where the users of a D2D link intend to communicate with one another while all the uplink/downlink cellular users coexist within the same shared spectrum. Using the power control strategy, Users A and B of D2D decide on the link's transmit power to prevent the throughput performance degradation of the coexisting uplink or downlink cellular link. Then, D2D users could attempt full duplex operation according to the controlled transmit power and its own self-interference cancellation capability.

\subsubsection{Uplink power-controlled full duplex cellular system}
Suppose that cellular users try to transmit an uplink signal, as shown in Fig.~\ref{fig:d2dcellular}-(b), to the BS while D2D users are connected with one another. Cellular Users A and B carry out uplink power control for successful D2D communications. Cellular users could also simultaneously transmit the uplink signal and receive the downlink signal depending on the controlled uplink transmit power and the possible self-interference cancellation level.

\subsubsection{Power-controlled full duplex D2D system underlying the downlink cellular network}
We also consider scenarios in which the user device concurrently deals with D2D and cellular signal  through full duplex operation. Figure~\ref{fig:d2dcellular}-(c) shows User A of a D2D underlying downlink cellular network. If the BS is operating in half duplex mode in a cellular network, the transmit power of D2D User A would have an effect on the downlink including that of User A. If the BS is operating in full duplex mode, the transmit power of D2D User A could also affect the uplink, including that of cellular User B, as well as the downlink. Depending on the duplexing mode of the cellular BS, D2D User A may  apply a power control strategy. In turn, D2D User A intends to transmit a D2D signal to D2D User B and, at the same time, receive a downlink signal from the BS. 

\subsubsection{Full duplex D2D system underlying the uplink power-controlled cellular network}
These scenarios are the cases where D2D User A tries to communicate with D2D User B while the D2D link shares the cellular uplink spectrum, as illustrated in Fig.~\ref{fig:d2dcellular}-(c). Via the uplink power control, D2D User~A could look for a chance to transmit a D2D signal to the D2D User B connecting with the cellular BS.

%%%%%%%%%%%%%%%%%%%%%%%%%%%%%%%%%
%%%%%%%%%%%%%%%%%%%%%%%%%%%%%%%%%
\subsection{System Specification \& Requirements}
As a starting point for designing suitable solutions to the aforementioned scenarios, which were introduced in the previous subsection, we need to be clear about the requirements for self-interference cancellation under a given system specification.

Between the input and output power, RF devices and circuits have nonlinear characteristics. For example, when input signal $\it{x}$ goes in an analog component like the power amplifier (PA) of a wireless transmitter, the PA creates outputs containing nonlinear cubic and higher order terms, such as $a{\it{x}}^{3}$, $b{\it{x}}^{5}$, and $c{\it{x}}^{7}$. The nonlinear properties of RF radios are a natural phenomenon that are mitigated by the effect of path loss. Because of this, a receiver in conventional wireless systems a receiver usually treats as nonlinear components from other transmitters. In a full duplex system, however, self-interference with high-powered nonlinear components has a dominant negative effect on performance without the compensation of generating nonlinear components~\cite{Sim2016}. To eliminate nonlinear components of self-interference, the authors in~\cite{Bharadia2013} proposed an approximation method of the nonlinear function using Taylor series expansion. The least-squares estimation method using parallel Hammerstein model was proposed in~\cite{Anttila2013, Anttila2014}. However, the solutions put forward in such previous works as~\cite{Bharadia2013,Anttila2013, Anttila2014} require of our scenarios onerous signal processing with high complexity.

Fig.~\ref{fig:Requirements} shows the requirements for self-interference cancellation under the given system specification. We focus on two analog elements -- a PA in the Tx chain and, in the Rx chain, another PA called low noise amplifier (LNA). These PAs play a central role in nonlinearity. With only a compact self-interference canceler adapted for operation with the user node, our key tactic to fulfill the performance requirements is to prevent, by visiting the nonlinear region of RF system, as much self-interference as possible. The first step is to investigate the range of transmit power to avoid or reduce the distortion of self-interference in the nonlinear region of Tx PA. The second step is to select the analog self-interference canceler, which directly makes the residual self-interference enter into the linear region of Rx LNA from a transmit power level at the user node, without needing signal processing with high complexity in the digital domain. 

As the input of PA continues to increase,  the gain at some point begins to decrease. The input power that causes the gain to drop 1 dB from the normal linear gain specification is called the {\it 1-{\rm{dB}} compression point}. And in PA specifications, this is used as a criterion for dividing linear and nonlinear operation. Let $P_{1 {\rm{dB}},T}$ and $P_{1 {\rm{dB}},R}$ denote 1-dB compression points of PA in the Tx chain and of LNA in the Rx chain, respectively. As we adopt orthogonal frequency division multiplexing (OFDM) transmission, defined in LTE standard for our prototype, we consider the special margin at the transmission side. Since OFDM waveform has an inherently high peak-to-average power ratio (PAPR), the relative difference between the highest and average transmit power is defined as the margin for PAPR at Tx side, denoted by $\Delta P_{m,T}$.  Thus, even though the transmit power is under 1-dB compression point of Tx PA, the nonlinearity could be exhibited depending on the margin for PAPR. The nonlinearity margin for PAPR is used to express the requirements of analog self-interference cancellation. Further, using this margin we present, in Section~\ref{sec:phyeval}, the impact of PA nonlinearity on the performance of our prototype in a real-world wireless channel. It could be a criterion for expecting the gain of full duplex operation, as a result of the scheduled power control at the user node. 

The high-powered input could saturate LNA and generate, as the output of LNA, a nonlinear component. Accordingly, self-interference should be sufficiently suppressed before LNA. We set the requirements of the minimum analog cancellation level and the maximum digital cancellation level, $\alpha_{min}$, $\delta_{max}$, as follows:
\begin{align}
\label{eq:minasic}
\alpha_{min} = \underbrace{P_{T} + \Delta P_{m,T}}_{\rm{highest\; transmit\;power}}-P_{1{\rm{dB}},R}
\end{align}

\begin{align}
\label{eq:maxdsic}
\delta_{max} = P_{T} + \Delta P_{m,T} - (P_{NF} + \alpha)
\end{align}
where $P_{T}$, $P_{NF}$, $\alpha$ are the power-controlled transmit power, the noise floor of the transceiver, and analog cancellation level, at the user node. The two sections that follow elaborate on the characteristics of the analog self-interference canceler. The canceler stably operates for the requirement~$\alpha_{min}$ with compactness and high power efficiency and 
the digital self-interference canceler design to attain as close to $\delta_{max}$ as possible.
%The two sections that follow elaborate on how to solve the key challenge and implement real-time full duplex radios.

%%%%%%%%%%%%%%%%%%%%%%%%%%%%%%%%%
%%%%%%%%%%%%%%%%%%%%%%%%%%%%%%%%%
\section{Full Duplex MIMO Front-End}
\label{sec:analogsic}

    \begin{figure}
    \centering
    \includegraphics[width = 3.5in]{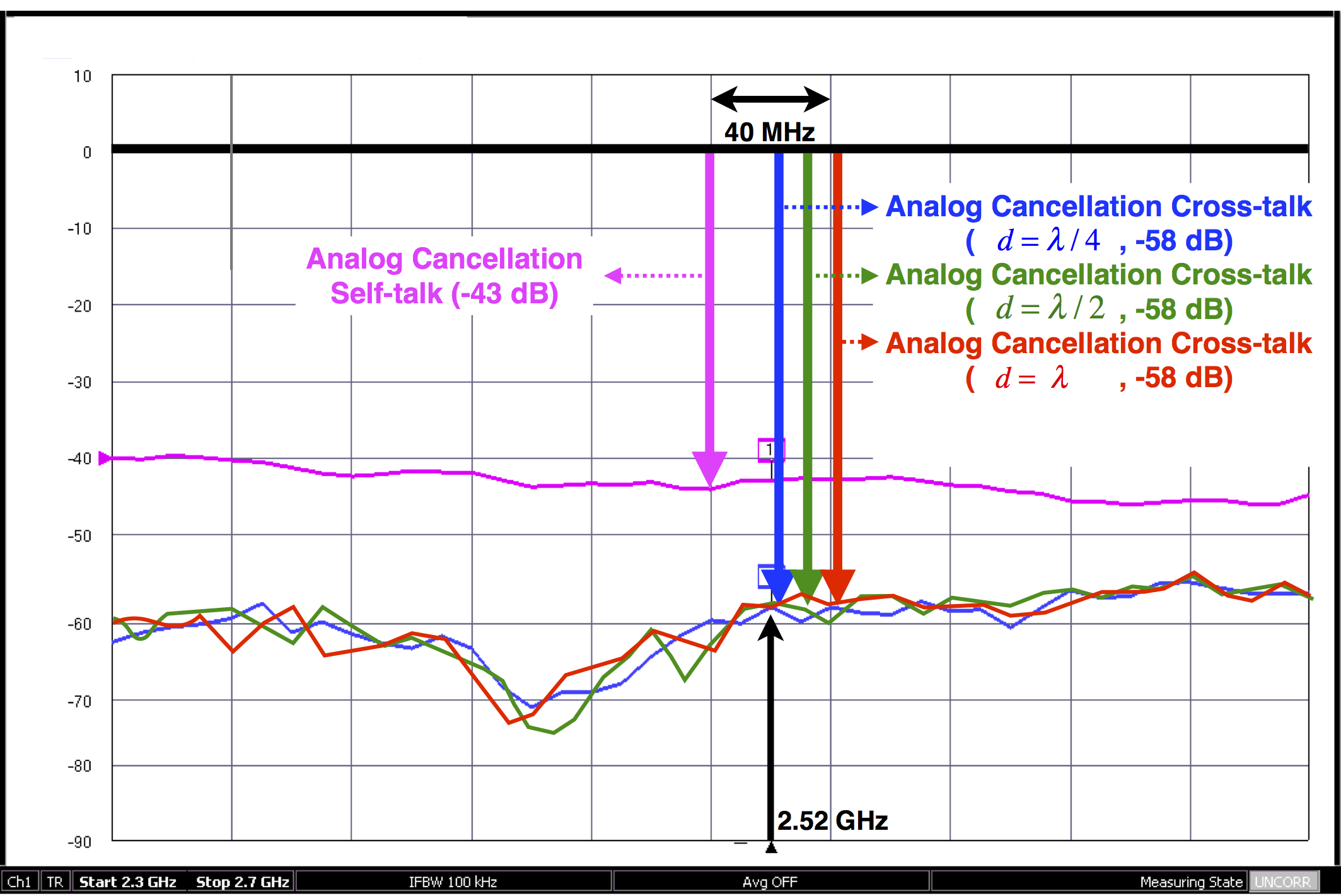}
    \caption{Measured isolation levels between Tx and Rx ports for self-talk (pink), cross-talk with an inter-antenna spacing $d=\lambda/4$ (blue), $\lambda/2$ (green), and $\lambda$ (red) with our dual-polarization MIMO antennas with the high XPD.}              
    \label{fig:ASIC}
    \end{figure}
In this section, we examine the characteristics of dual-polarized MIMO RF front-end as the analog self-interference canceler. In conventional uni-directional communications, for a diversity gain of MIMO channels, an inter-antenna spacing on the order of $\lambda/2$ is required to reduce the correlation between the signals on different channels. Such wide spacing leads to an impractical MIMO architecture for the mobile user. The authors in~\cite{Beza2015, Anreddy2006} presented the dual-polarized antenna as an attractive alternative for realizing compact MIMO architectures. On the basis of the advantage, we explore the benefits of exploiting the dual-polarized antenna as it pertains to compact full duplex MIMO architecture. 

  \begin{figure*}
    \centering
    \includegraphics[width = 7.2in]{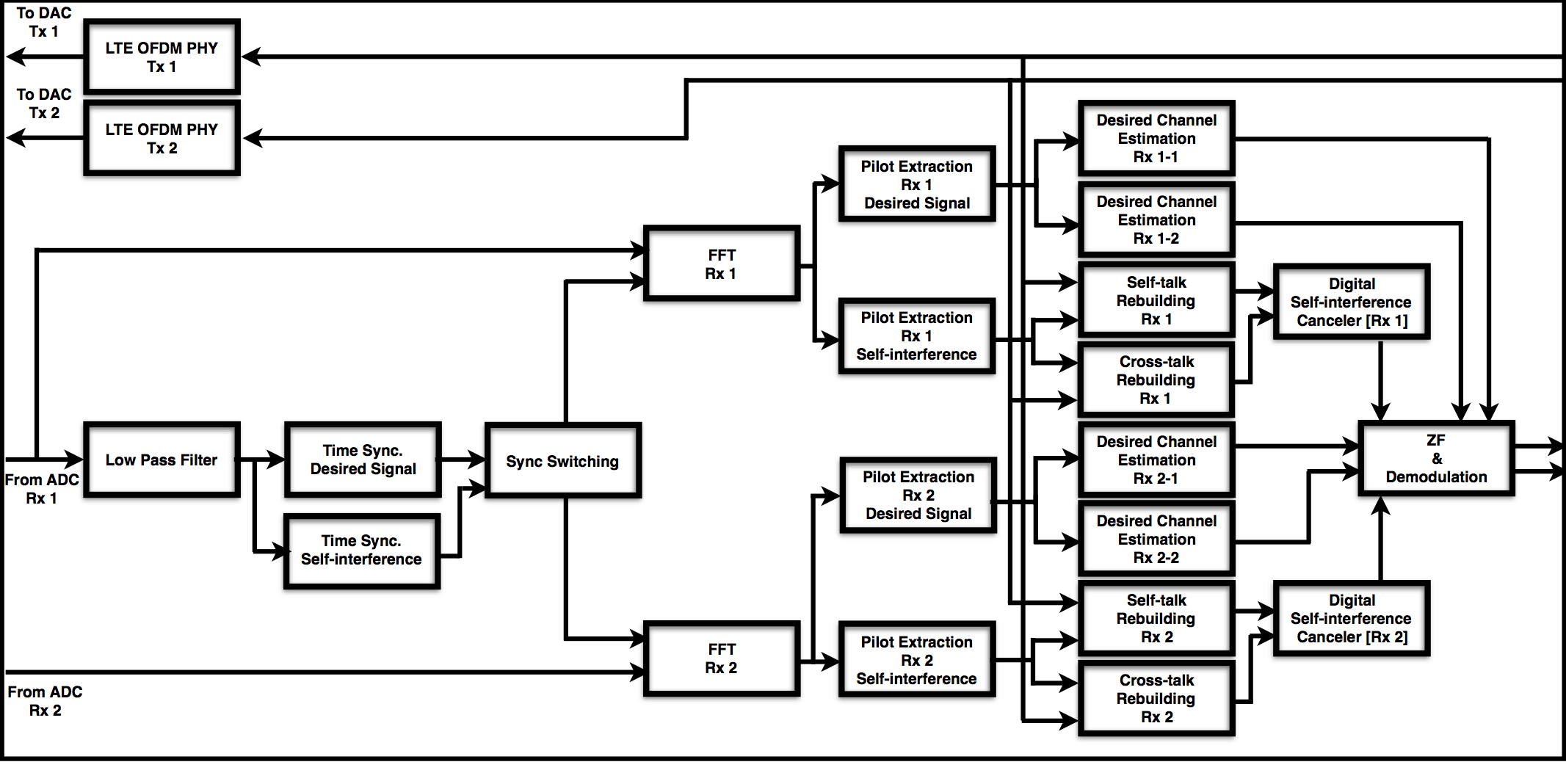}
    \caption{Block diagram of full duplex MIMO PHY design on multiple FPGA modules. We jointly implement digital self-interference canceler on LTE-based MIMO OFDM PHY including synchronizer, channel estimator, and demodulator.}              
    \label{fig:DSIC}
    \end{figure*}

As MIMO expands from a SISO configuration in a full duplex radio, it incurs self-interference sources between inter-antennas, in its own radio, as well as self-interference generated at its own antenna. Owing to this phenomenon, the expansion to an $N$-dimensional full duplex system creates a number of $N^2$ self-interference sources. For the sake of clarity, we define self-interference generated at own antenna is defined as {\it{self-talk}} and self-interference generated between inter-antennas in a full duplex node as {\it{cross-talk}}. In our full duplex~$2 \times 2$~configuration, self-interference sources make up, as illustrated in Fig.~\ref{fig:MIMO_FD}, two self-talks and two cross-talks. These self-interference terms are passively suppressed though a dual-polarized RF front end with a high XPD  in the analog domain, without a copy of self-interference by additional RF chains. The XPD is defined as the ratio of the co-polarized average received power $P_{\parallel}$ to the cross-polarized average received power $P_{\bot}$. Thus, high XPD indicates an outstanding ability not to radiate into orthogonal cross-polarization from co-polarized transmission, and vice versa~\cite{oh2014}. In this regard, co-pole and cross-pole  are used for a dedicated transmission and reception path, to construct in our case a single shared antenna between the Tx and Rx ports. Due to the high XPD characteristic, our analog self-interference canceler could achieve high isolation between Tx and Rx ports using only a passive strategy. 

For investigation of the isolation level between Tx and Rx ports, the S-parameter measurements of dual-polarization MIMO RF were performed in an indoor environment using Anritsu MS4647A Vector Network Analyzer. As can be seen in Fig.~\ref{fig:ASIC}, the isolation level for the self-talk achieved approximately -43~dB. This numerical value indicates the analog self-interference cancellation level through a passive strategy, when the single dual-polarized antenna is employed in a full duplex SISO radio. We measured the isolation levels between ports of co-pole at RF~1 and cross-pole at RF~2, and the co-pole at RF~2 and cross-pole at RF~1, as illustrated in Fig.~\ref{fig:MIMO_FD}. We did so to characterize cross-talk suppression, with an inter-antenna spacing of~$d\in \{\lambda/4, \lambda/2, \lambda \}$. One remarkable point is that the isolation level for cross-talk is still maintained with approximately -58~dB in spite of decreasing the inter-antenna spacing from $\lambda$ to $\lambda/4$. In previous work on full duplex MIMO structure, cross-talk suppression, via the passive strategy, has depended on the inter-antenna spacing. This requires a bulky MIMO structure or active strategies with high complexity for residual cross-talk cancellation.
%within the range of the inter-antenna spacing from $\lambda/4$ to $ \lambda$.

The residual interference power after analog cancellation~$P_{A}$ is expressed in terms of the analog self-talk cancellation level~$\alpha_{s}$ and analog cross-talk cancellation level~$\alpha_{c}$ as follows.
\begin{align}
P_{A} = \underbrace{  {\frac{1}{2}}P_{T} \times 10^{(\alpha_{s}/10)} }_{\rm{residual\;self-talk\;power}}+
\underbrace{  {\frac{1}{2}}P_{T} \times 10^{(\alpha_{c}/10)} }_{\rm{residual\;cross-talk\;power}}.
\end{align}
The total analog cancellation level $\alpha$ is defined as the ratio of  residual interference power after analog cancellation $P_{A}$ to power-controlled transmit power $P_{T}$. From the measurement results, i.e. $\alpha_{s}=-43$, $\alpha_{c}=-58$, we could confirm that the total analog cancellation level $\alpha$ of approximately -46~dB was achieved through the passive method only.

%%%%%%%%%%%%%%%%%%%%%%%%%%%%%%%%%
%%%%%%%%%%%%%%%%%%%%%%%%%%%%%%%%%
\section{Full Duplex MIMO Physical Layer}
\label{sec:fullmimophy}

  \begin{figure}
    \centering
    \includegraphics[width = 3.5in]{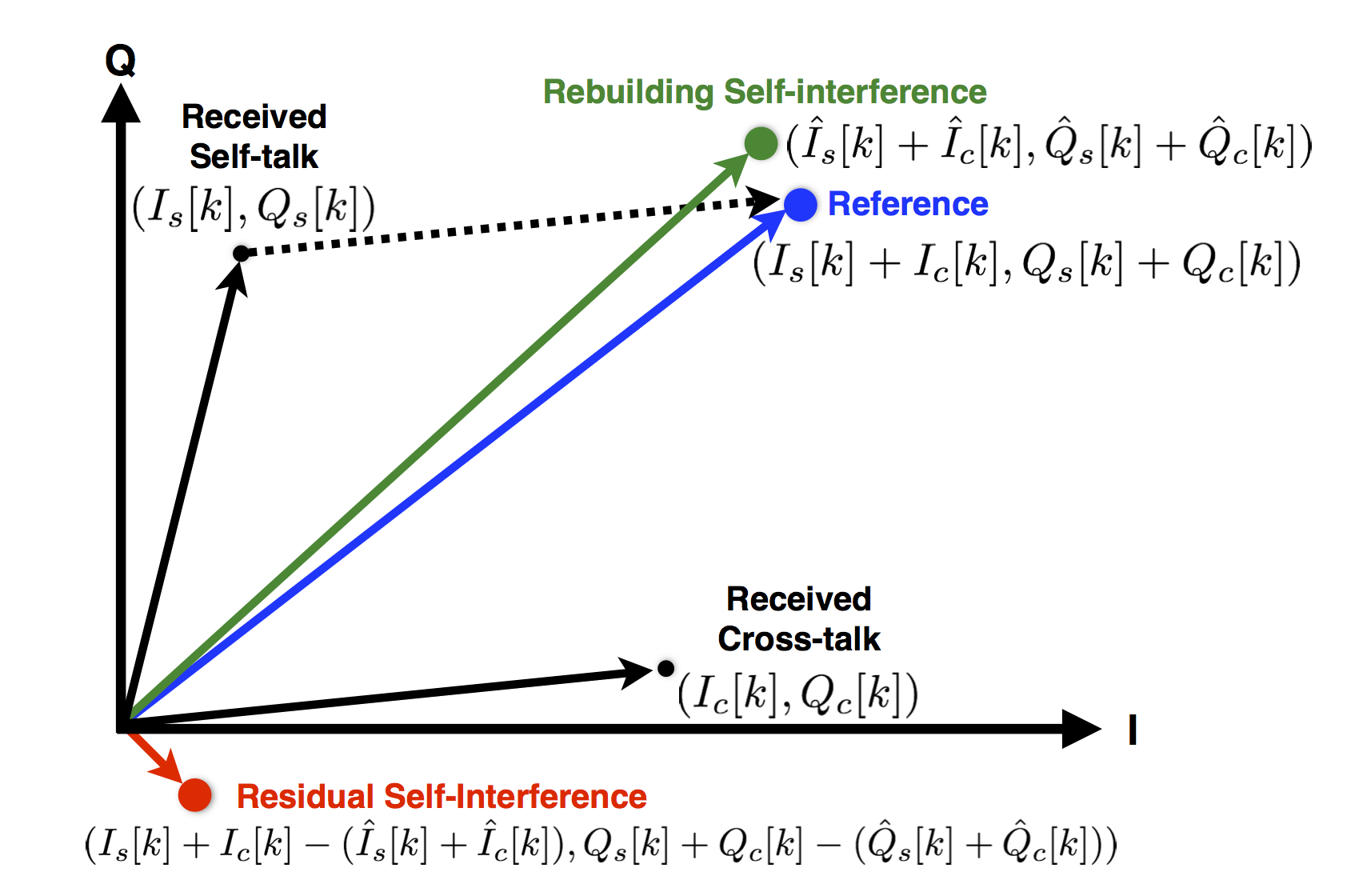}
    \caption{Metric for digital self-interference cancellation level. The vector sum of received self-talk and cross-talk could be considered the reference vector (blue) of digital self-interference cancellation level. The error vector between the reference and the rebuilding self-interference vector (green)  is a residual self-interference vector (red) in full duplex MIMO PHY. By these vectors, we could calculate how much self-interference is mitigated from the original interference sources in the baseband through the digital self-interference canceler. }              
    \label{fig:METRIC}
    \end{figure}

    \begin{figure*}[t]
    \centering
    \includegraphics[width = 7.2in]{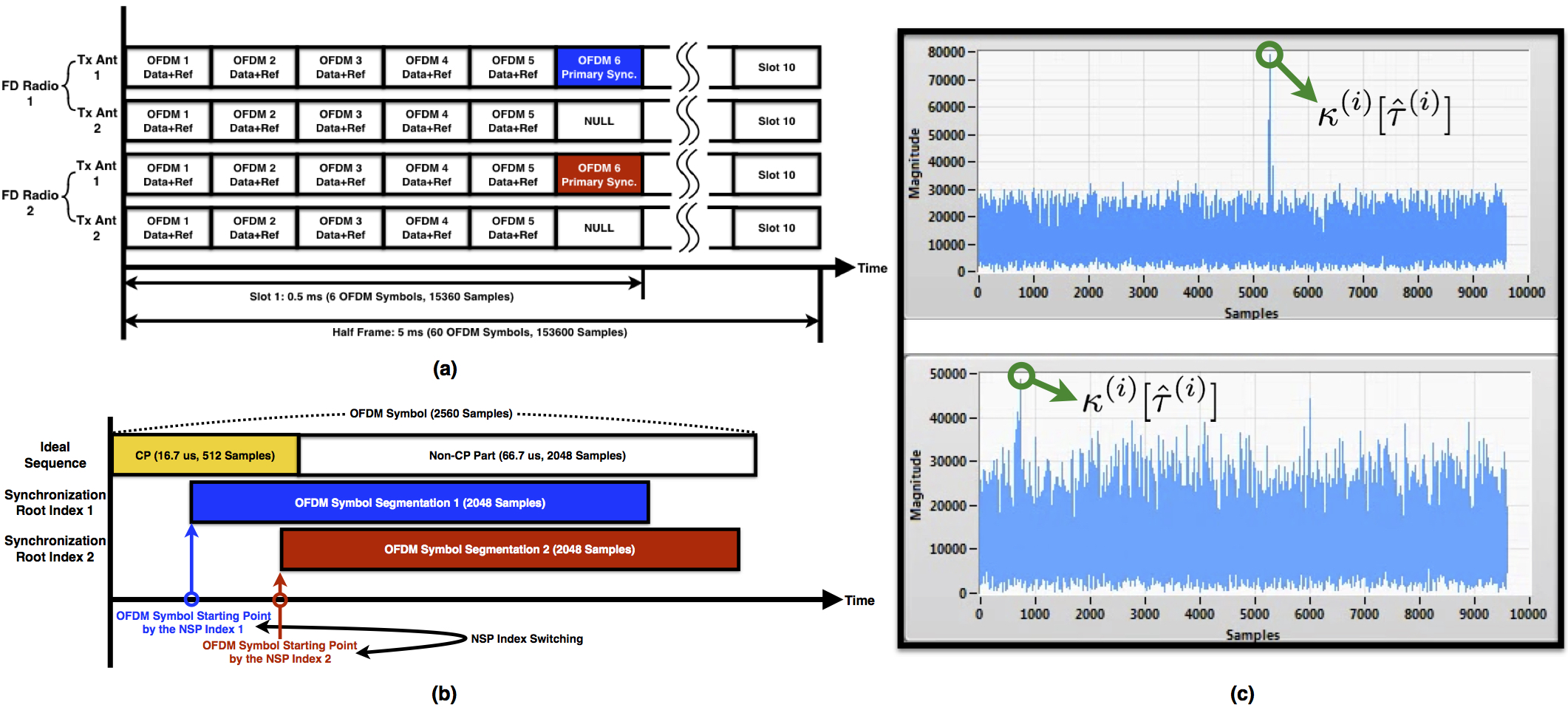}
    \caption{Synchronization protocol for full duplex $2 \times 2$ MIMO communications. Frame structure including PSS in our prototype is modified from the conventional LTE frame structure to sustain compatibility with half duplex mode in our scenarios as well as the comparison with the actual performance of conventional half duplex LTE system. (a) Frame structure for a full duplex $2 \times 2$ MIMO link (b) the NSP index switching for full duplex synchronization. If the difference of over-the-air propagation delays between the desired signal and self-interference at a full duplex node is within CP duration (16.7 $\mu s$), OFDM symbols could be decoded by  OFDM symbol segmentation using any information of the NSP indices of root index 1 and 2. (c) Comparison of NSP peak values depending on the distance or channel quality between a full duplex link.}              
    \label{fig:Sync_Ref}
    \end{figure*}
    
In this section, we describe the core ideas of the full duplex MIMO PHY architecture. Figure~\ref{fig:DSIC} shows a block diagram of our proposed full duplex MIMO PHY. We first introduce the performance metric for digital self-interference cancellation.

%%%%%%%%%%%%%%%%%%%%%%%%%%%%%%%%%
%%%%%%%%%%%%%%%%%%%%%%%%%%%%%%%%%
\subsection{Metric}
\label{subsec:metric}
In the analog domain, the performance metric of self-interference suppression is the isolation level between the Tx and Rx ports by S--parameter measurements. In the baseband, the performance metric of digital self-interference cancellation is the \textit{modified error vector magnitude}. The error vector magnitude, commonly used to quantify the performance of a digital radio transceiver, is defined as the ratio of the reference power to the error vector power that exists between the original constellation point and its received one at the receiver. In this definition, the reference indicates the original constellation point. The digital self-interference cancellation level is defined as the ratio of the power of residual self-interference after digital cancellation to the power of the received self-interference in the baseband. We calculate the digital self-interference cancellation level by measuring of the residual interference vector and the actual sum vector of self-talk and cross-talk in the I/Q plane, as shown in Fig.~\ref{fig:METRIC}.
$e_{k}$ denotes the power of the error vector magnitude between the reference and rebuilding self-interference of $k$th sample:
\begin{align}
\begin{split}
e_{k}=&\{I_{s}[k]+I_{c}[k]-(\hat I_{s}[k]+\hat I_{c}[k])\}^2\\
&+\{Q_{s}[k]+Q_{c}[k]-(\hat Q_{s}[k]+\hat Q_{c}[k])\}^2
\end{split}
\end{align}
where  $I_{s}[k]$, $I_{c}[k]$, $\hat I_{s}[k]$, $\hat I_{c}[k]$, respectively, represent the in-phase component of the received self-talk, cross-talk, and rebuilding self-talk, cross-talk, respectively. And $Q_{s}[k]$, $Q_{c}[k]$, $\hat Q_{s}[k]$, $\hat Q_{c}[k]$ represent the quadrature component of the received self-talk, cross-talk and rebuilding self-talk, cross-talk, respectively. Applying $e_{k}$, the instantaneous digital self-interference cancellation level of the $k$th~sample is given by 
\begin{align}
\delta_{k}=
10 \; {\rm log}_{10}
\left[ {
\frac
{e_{k}}{(I_{s}[k]+I_{c}[k])^2+(Q_{s}[k]+Q_{c}[k])^2} 
} \right]
\end{align}

%%%%%%%%%%%%%%%%%%%%%%%%%%%%%%%%%
%%%%%%%%%%%%%%%%%%%%%%%%%%%%%%%%%
\subsection{Signal Model}
\label{subsec:sigmod}
Consider a two-node full duplex channel where each full duplex transceiver with $N_{t}$ Tx antennas and $N_{r}$ Rx antennas simultaneously sends and receives $N_{s}$ independent streams. After a fast Fourier transform (FFT) operation of OFDM symbols at the Rx block, the input-output relationship for the subcarrier index $k$ at each node is given by
\begin{align}
\label{eq:signalmodel}
\textbf{y}[k]=
\textbf{H}[k]{\textbf{x}_{D}}[k]+\textbf{G}[k]{\textbf{x}_{S}}[k]+
{\textbf{z}}[k],
\end{align}
where $\textbf{x}_{D}[k], \textbf{x}_{S}[k]\in {\mathbb{C}}^{{N_{t}} \times 1}$ denote the desired symbol vector sent by its associated node and the self-interference symbol vector generated from its own node, respectively. $\textbf{H}[k], \textbf{G}[k]~\in~{\mathbb{C}}^{N_{r} \times N_{t}}$ represent the desired channel matrix from its associated node and the self-interference channel matrix from its own node. $\textbf z[k]$ denotes the noise vector in the system. We focus on the case of $N_{t}=2, N_{r}=2$, and $N_{s}=2$ to explain the algorithm and structure of full duplex MIMO PHY based on our implementation.

%%%%%%%%%%%%%%%%%%%%%%%%%%%%%%%%%
%%%%%%%%%%%%%%%%%%%%%%%%%%%%%%%%%
\subsection{Proposed Synchronization Design}
\label{subsec:syncdesign}

\begin{table*}[t] 
\begin{center}
\caption{Proposed synchronization algorithm for full duplex.} 
\renewcommand{\arraystretch}{1.2}
\label{tb:sync_algorithm}
\begin{tabular}{c|c}
\hline
\bfseries{Step} & \bfseries{Algorithm}\\ 
\hline\hline
Step 1 & Operate low-pass filtering using the received signal by the first Rx antenna\\
Step 2 & Calculate cross-correlation  $\kappa^{(1)}[d], \kappa^{(2)}[d]$\\
Step 3 & Calculate $\alpha^{(1)}, \alpha^{(2)}$ such that $\hat\tau^{(1)}=\arg\max_{d} \kappa^{(1)}[d]$, $\hat\tau^{(2)}=\arg\max_{d} \kappa^{(2)}[d]$ \\
Step 4  & Test Feasibility condition $\alpha^{(1)} > \alpha_{th}$ or $\alpha^{(2)} > \alpha_{th}$. If this condition is satisfied, go to Step 5. Otherwise, go to Step 2.  \\
Step 5 & 
Update $\tau^{(1)}, \tau^{(2)}$ such that 
	$\left\{ \displaystyle\begin{array}{ll}
		\displaystyle \tau^{(1)}=\hat\tau^{(1)}, \tau^{(2)}=\hat\tau^{(2)}, & \mbox{if $\alpha^{(1)}>\alpha_{th}$ and  $\alpha^{(2)}>\alpha_{th}$}\\
		\displaystyle \tau^{(1)}=\hat\tau^{(1)}, \tau^{(2)}=\hat\tau^{(1)}, & \mbox{if $\alpha^{(1)}>\alpha_{th}$ and  $\alpha^{(2)} \leq\alpha_{th}$}\\
		\displaystyle \tau^{(1)}=\hat\tau^{(2)}, \tau^{(2)}=\hat\tau^{(2)}, & \mbox{if $\alpha^{(1)} \leq \alpha_{th}$ and  $\alpha^{(2)}>\alpha_{th}$}\\
	\end{array}
	\right.
	.$\\
\hline
\end{tabular}
\renewcommand{\arraystretch}{1}
\end{center}
%\hfill\footnotesize{$^a$ Possible only under specific conditions.}
\end{table*}

A key requirement to initiate full duplex communications between scheduled nodes is the synchronization protocol. Furthermore, to stabilize the total self-interference cancellation performance from the analog to the digital domain, it is imperative to design a robust synchronization algorithm.
%Furthermore, to measure the actual performance, i.e., total self-interference cancellation level with both of analog and digital domain, link throughput in real-world wireless channel, of the full duplex link, full duplex implementation with suitable synchronization protocol is imperative.

We modify the frame structure of the LTE downlink with a frame duration of 10 ms for transmission~\cite{sesia2009lte}. As illustrated in Fig.~\ref{fig:Sync_Ref}-(a), each slot includes 6 OFDM symbols with 512 cyclic prefix (CP) lengths (extended mode in LTE). The primary synchronization symbol (PSS) is situated in the last OFDM symbol of the first slot of each half frame at the first Tx antenna of each full duplex node. The PSS is modulated on subcarrier index $k$, around the DC-carrier, by a length-62 Zadoff-Chu sequence given as,
\begin{align}
\label{eq:pss}
	P_{i}[k] = 
    \left\{ 
    	\begin{array}{ll}
    	\displaystyle{e^{ - j\frac{\pi }{N_{p}}u^{(i)}k\left( {k + 1} \right)}} & \text{if } - 31 \le k \le - 1\\
        \vspace{0.01cm}    &  \\
        \displaystyle{e^{ - j\frac{\pi }{N_{p}}u^{(i)}\left( {k + 1} \right)\left( {k + 2} \right)}} & \text{if } 1 \le k \le 31
        \end{array}
    \right.,
\end{align}
where $u^{(i)}$ is an element among the root indices defined in LTE standard, and $N_{p}$ is the sequence length for the PSS. Exploiting the characteristic, where the Zadoff-Chu with a different root index is orthogonal to each other, we apply two root indices, i.e., $u^{(1)}=25$, $u^{(2)}=29$, to detect time offset of the self-interference as well as the desired signal, by over-the-air propagation delays at each full duplex receiver. Further, using the knowledge regarding the two kinds of time offset information, we propose an algorithm to enhance the synchronization probability between full duplex nodes. 

Since the PSS is allocated in a part  around the DC-carrier of its associated OFDM symbol, a low-pass filtering of baseband OFDM symbol helps to extract the only PSS subcarriers from the received OFDM symbol containing data subcarriers as well. The low-pass filtering is operated as a reference signal for timing synchronization using the received signal by the first Rx antenna at each node. The filtered signal at the $n$th time sample is given by,
\begin{align}
\label{eq:LPF}
{y_{f}}[n]=\sum_{\ell=0}^{N_{f}}  \sum_{j=1}^{2} \sum_{i=1}^{2} y_{1,j}^{(i)}[n-\ell]f[\ell]\end{align}
where $f[\ell]$ represents an impulse response of finite length $N_{f}$ for $\ell \in \{{1,2,...,N_{f}}\}$. If $i=1$, $y_{1,j}^{(i)}$ represents the received signal from the partner's $j$th Tx antenna to the first Rx antenna; if $i$=2, the received signal from own transmitter's $j$th antenna to the first Rx antenna for $j \in \{1,2\}$. Each node has ideal Zadoff-Chu sequences for two kinds of root indices -- as a synchronization code for desired signal and self-interference, given as,
\begin{align}
\label{eq:PSSIDFT}
p^{(i)}[n]=\sum_{k=-{N_{p}/2}}^{N_{p}/2} P_{i}[k]e^{j{\frac{2\pi}{N}}kn}
\end{align}
which is an $N$-point inverse discrete Fourier transform (IDFT) of a Zadoff-Chu sequence, having root index $u^{(i)}$, in~(\ref{eq:pss}). In order to detect the starting point of the reference OFDM symbol, we find the index of the maximum value of cross-correlation with the ideal sequence $p^{(i)}[n]$ with respect to root index $u^{(i)}$ and the received sequence after low-pass filtering $y_{f}[n]$ given as,
\begin{align}
\label{eq:Cross}
\kappa^{(i)}[d]= \left| {{\sum_{n=0}^{N-1}  {y_{f}}[n+d] p^{(i)*}[n]}} \right|
\end{align}

\begin{align}
\label{eq:PIndex}
\hat\tau^{(i)}=\arg\max_{d} \kappa^{(i)}[d]
\end{align}
where $x^*$ denotes the complex conjugate of complex value $x$, and $d$ represents the lag between two sequences.

Let us define a parameter for the ratio between the peak and average values of the cross-correlation in (\ref{eq:Cross}). We call this normalized synchronization peak (NSP) value:
\begin{align}
\label{eq:alpha}
\alpha^{(i)}=\frac{\kappa^{(i)}[\hat\tau^{(i)}]}{ \frac{1}{N_{c}} \sum_{d=0}^{N_{c}}\kappa^{(i)}[d]}
\end{align}
To decide the timing synchronization, we set the threshold value denoted by $\alpha_{th}$ based on the conventional LTE system. For example, we consider a case where $\alpha^{(i)}>\alpha_{th}$ as the peak index detection for timing synchronization is performed successfully. 

As shown in Fig~\ref{fig:Sync_Ref}-(b), it is assumed that the difference in over-the-air propagation delays between the desired signal and the self-interference signal is within CP duration (16.7~$\mu s$). This assumption is valid in our indoor experimental setup because the distance between a full duplex link is not the source of an OFDM symbol-level timing difference between the desired signal and the self-interference signal. For this reason, one might argue that it is possible to achieve enough timing synchronization between the full duplex link using only some of the NSP index information for the desired signal and self-interference. However, as the desired and self-interference signals affect each other as strong interference, the mutual feasibility condition for synchronization cannot be satisfied. For example, in the case where full duplex nodes are near one another, the desired signal-- relatively stronger than the self-interference at its own node due to analog cancellation-- lowers the NSP value for the synchronization of each self-interference. In contrast,  in the case where the two nodes are far apart, self-interference at each node-- stronger than the desired signal-- drops the NSP value for synchronization of the own desired signal. 

To solve this problem, we propose a novel method to improve the probability of successful synchronization in a full duplex link. In Section~\ref{sec:phyeval}, we show the implementation results. The basic idea of the proposed synchronization algorithm is that it makes good use of the NSP indices through two kinds of PSSs, which consist of different root indices,  transmitted from each full duplex. As the possible cases from the feasibility condition test result for each NSP value, it updates the NSP indices for synchronization of the desired signal and the self-interference. The update method of the NSP index is as follows.

\subsubsection{Case of $\alpha^{(1)}>\alpha_{th}$, $\alpha^{(2)}>\alpha_{th}$}
When the degradation level of the desired signal by path loss is on the order of the analog self-interference cancellation level, then both NSP values generated from own radio and its full duplex partner are likely to satisfy each feasibility condition. Therefore, it updates each calculated NSP index, given as,
\begin{align}
\tau^{(1)}=\hat\tau^{(1)}, \tau^{(2)}=\hat\tau^{(2)}
\end{align}
where $\tau^{(1)}$ and $\tau^{(2)}$ represent the updated NSP index for the desired signal and for the self-interference, respectively.

\subsubsection{Case of $\alpha^{(1)}>\alpha_{th}$, $\alpha^{(2)} \leq\alpha_{th}$}
The power of the residual self-interference gets through the analog canceler and the analog-to-digital converter (ADC). When such power is much smaller than that of the desired signal sent from its partner in close proximity, we should exploit the only NSP index by PSS sent from the partner. This enables us to synchronize self-interference as well as the desired signal, given as,
\begin{align}
\tau^{(1)}=\hat\tau^{(1)}, \tau^{(2)}=\hat\tau^{(1)}
\end{align}

\subsubsection{Case of $\alpha^{(1)} \leq \alpha_{th}$, $\alpha^{(2)}>\alpha_{th}$}
When the channel condition between a full duplex link is of poor quality, the PSS sent from own transceiver is more useful at detecting the NSP index than is the PSS sent from its partner. In this case, the desired signal and self-interference is synchronized based on the NSP index by self-interference, given as, 
\begin{align}
\tau^{(1)}=\hat\tau^{(2)}, \tau^{(2)}=\hat\tau^{(2)}
\end{align}
Table~\ref{tb:sync_algorithm} summarizes the proposed NSP-switching algorithm for synchronization between full duplex nodes.

%%%%%%%%%%%%%%%%%%%%%%%%%%%%%%%%%
%%%%%%%%%%%%%%%%%%%%%%%%%%%%%%%%%
\subsection{Proposed Digital Self-Interference Canceler}
\label{subsec:dsic}
What stands out regarding our digital self-interference canceler implementation is its novel architecture. Its contributions over all prior full duplex implementations consist of two significant ones: 1) high resistibility to frequency-selective fading and 2) high compatibility with conventional LTE systems. For network scenarios where the wireless channel is frequency-selective and users demand broadband service, we consider that the per-subcarrier digital cancellation method is more effective than the digital cancellation in the time domain. Therefore, we focus on the per-subcarrier digital canceler with low computational complexity based on LTE MIMO PHY structure.

 \begin{figure}[!t]
 \centering
 \includegraphics[width = 3.5in]{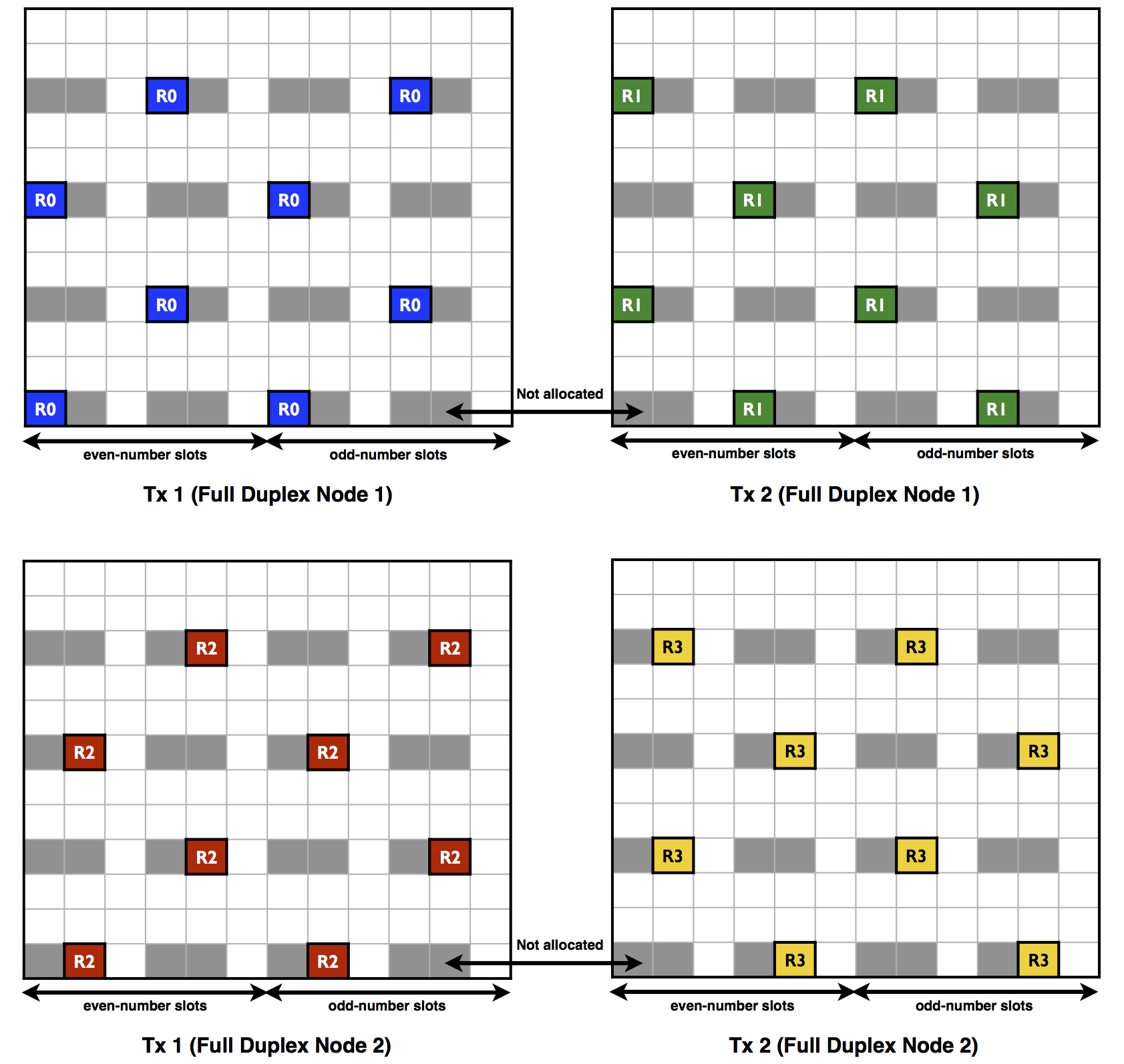}
\caption{Reference symbol allocation for a full duplex $2 \times 2$ MIMO link. }              
\label{fig:REF}
\end{figure}

\begin{figure*}[t]
\centering
\includegraphics[width = 7.2in]{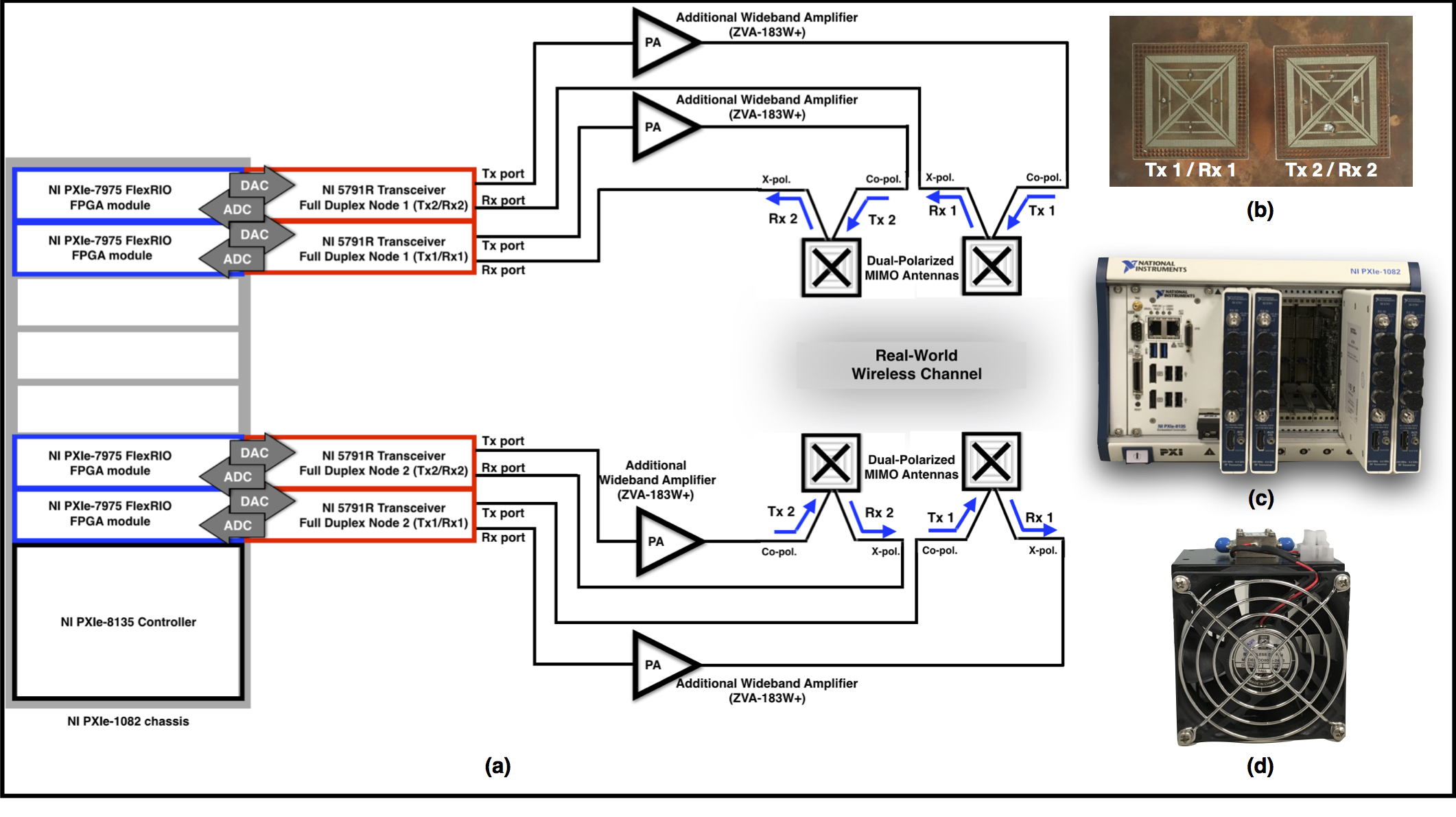}
\caption{Hardware configurations of compact full duplex MIMO radios: (a) physical layout of a full duplex $2 \times 2$ MIMO link. (b) dual-polarized full duplex MIMO front-end. (c) SDR platform for implementation and operation of full duplex MIMO PHY (d) additional wideband amplifier (ZVA-183W+) for transmission using the maximum Tx power of cellular of D2D user.}              
\label{fig:Hardware}
\end{figure*}
To suppress any residual self-interference in a baseband, full duplex PHY receiver, there are, in addition to the conventional architecture, three things that are necessary-- the modules for self-interference channel estimation, self-interference rebuilding, and its subtraction from the received signal. Our PHY prototype manages  all of these processes after FFT block at the Rx side for the per-subcarrier cancellation strategy, as depicted in Fig.~\ref{fig:DSIC}. For channel estimation of a full duplex link, which has a total of four Tx antenna, we adopt a cell-specific reference signal (CRS) pattern for a four-antenna port configuration in LTE, as illustrated in Fig.~\ref{fig:REF}. Exploiting CRS, each full duplex node could simultaneously estimate a total of eight wireless channels including four self-interference channels from its own radio. From the signal model in (\ref{eq:signalmodel}), the estimated channel for the subcarrier index $k$ at each node is expressed by 
\begin{align}
\label{eq:estimateM}
\hat{\textbf{H}}[k] = \begin{bmatrix} \hat h_{1,1}[k]&\hat h_{1,2}[k]\\ \hat h_{2,1}[k]&\hat h_{2,2}[k] \end{bmatrix},\;
\hat{\textbf{G}}[k] = \begin{bmatrix} \hat g_{1,1}[k]&\hat g_{1,2}[k]\\ \hat g_{2,1}[k]&\hat g_{2,2}[k] \end{bmatrix}
\end{align}
where $\hat h_{i,j}[k]$ represents the estimate of the desired channel coefficient from partner's Tx port $j$ to own Rx port $i$, operating in full duplex mode. $\hat g_{i,j}[k]$ represents the estimate of the self-interference channel coefficient from Tx port $j$ to Rx port $i$, i.e., if $i=j$, $\hat g_{i,j}[k]$ is the estimate of self-talk channel coefficient at antenna $i$ or $j$, if $i \neq j$, $\hat g_{i,j}[k]$ is the one of cross-talk channel from antenna $j$ to $i$. Using the estimated result of the self-interference channel coefficient per subcarrier, self-interference rebuilding operates in the serial order of the subcarrier index $k$ given as, 
\begin{align}
\begin{split}
\hat y_{S}[k]=\underbrace{\hat g_{1,1}[k]x_{S,1}[k]+\hat g_{2,2}[k]x_{S,2}[k]}_{\rm rebuilt\;self-talk}\\
+\underbrace{\hat g_{2,1}[k]x_{S,1}[k]+\hat g_{1,2}[k]x_{S,2}[k]}_{\rm rebuilt\;cross-talk}
\end{split}
\end{align}
where $x_{S,i}$ denotes the independent message sent from the $i$th Tx antenna of its own node to its partner. At each Rx port of a full duplex $ 2 \times 2$ radio, one self-talk and cross-talk act as a residual interference source in the baseband. Therefore, per-subcarrier digital self-interference cancellation is mathematically expressed by
\begin{align}
\begin{split}
\textbf{y}[k]-\hat {\textbf{y}}_{S}[k]&=
\begin{bmatrix}
y_{1}[k]-\hat y_{S,1}[k]\\
y_{2}[k]-\hat y_{S,2}[k]
\end{bmatrix}\\
&=\begin{bmatrix}
y_{1}[k]-(\underbrace{\hat g_{1,1}[k]x_{S,1}[k]+\hat g_{1,2}[k]x_{S,2}[k]}_{\rm rebuilt\;sample\;for\;Rx\;port\;1})\\
y_{2}[k]-(\underbrace{\hat g_{2,1}[k]x_{S,1}[k]+\hat g_{2,2}[k]x_{S,2}[k]}_{\rm rebuilt\;sample\;for\;Rx\;port\;2})
\end{bmatrix}
\end{split}
\end{align}
where $y_{i}[k]$, $\hat y_{S,i}[k]$ represent the $k$th received subcarrier sample and a sum of rebuilt samples for residual self-interference sources, respectively, at the $i$th Rx port. Applying the performance metric in Section~\ref{subsec:metric}, we measured the instantaneous digital self-interference cancellation level in our SDR platform in various indoor environments. It was found that the maximum digital self-interference cancellation level $\delta_{max}$ achieved approximately $\text {-55}$~dB through our implemented real-time digital canceler. Results are detailed further in Section~\ref{sec:phyeval}.

\begin{figure*}[t]

\centering
\includegraphics[width = 7.2in]{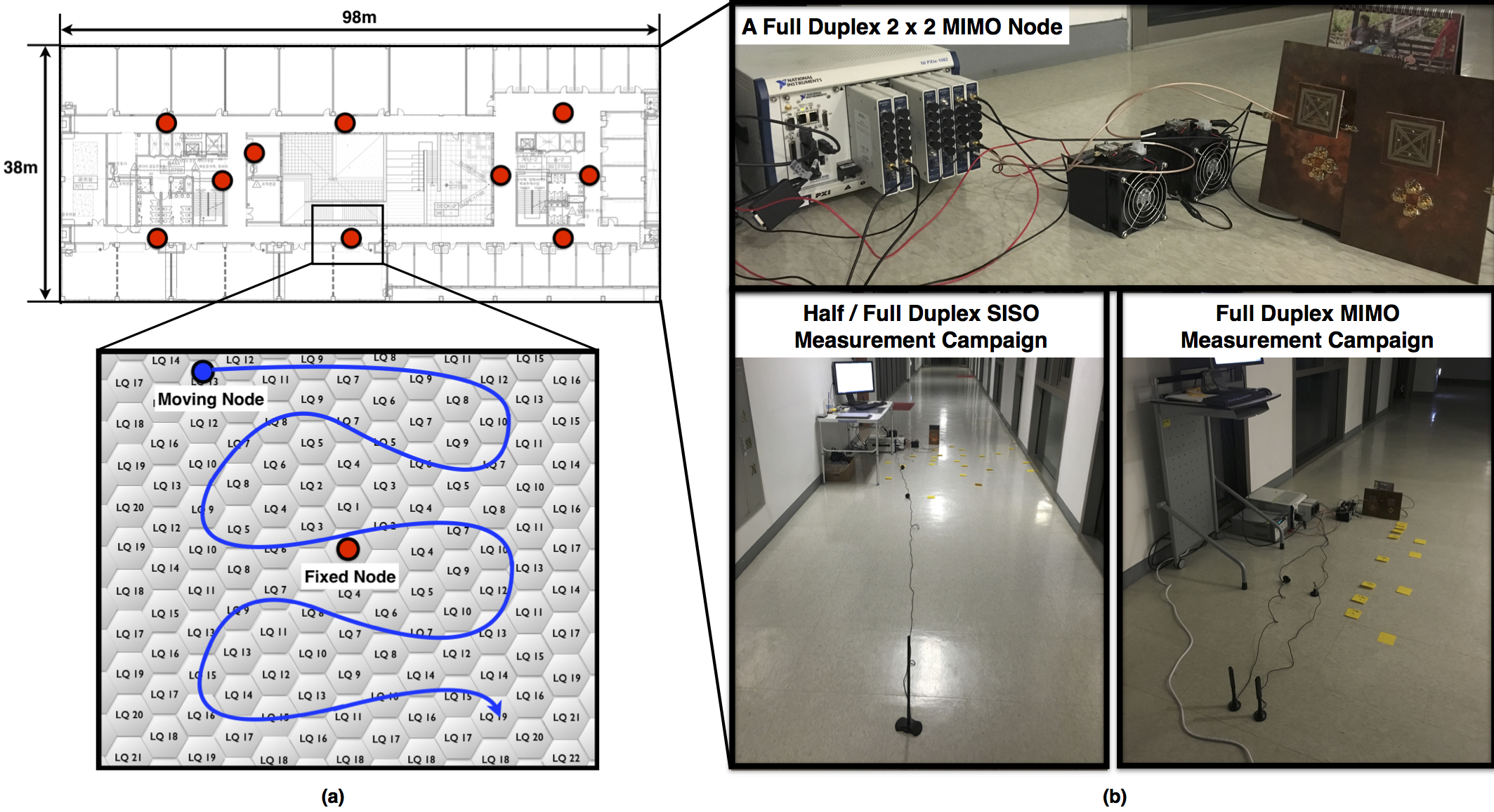}
 \caption{Experimental scenario and setup in indoor. (a) measurement locations (red circle) in the building and link quality map regarding real-world indoor network (b) physical setup of our prototypes in an indoor testbed.}
\label{fig:Expscen}              
 \end{figure*}

To evaluate the performance of a full duplex MIMO link, we implement a MIMO decoder to extract two independent messages. 
\begin{align}
\hat{\textbf{x}}_{D}[k]=
\begin{bmatrix}
\hat x_{D,1}[k]\\
\hat x_{D,2}[k]
\end{bmatrix}={\textbf{F}}[k]
(\textbf{y}[k]-\hat {\textbf{y}}_{S}[k])
\end{align}
where $\hat{\textbf{x}}_{D}[k]$ denotes the decoded symbol vector, which has two independent decoded symbols $\hat x_{D,1}[k], \hat x_{D,2}[k]$ sent from the associated partner. $\textbf{F}[k]$ represents the MIMO decoding matrix. We apply a zero-forcing MIMO detector using the desired channel matrix's estimate, ${\hat{\textbf{H}}}[k]$, in (\ref{eq:estimateM}) given as,
\begin{align}
\textbf{F}[k]=
{({{\hat{\textbf{H}}}[k]^{H}}{\hat{\textbf{H}}}[k])^{-1}}{{\hat{\textbf{H}}}[k]^{H}}.
\end{align}

%To operate our compact full duplex MIMO radio in real-time in a changing environment, even perform the performance characterization, proposed full duplex MIMO PHY architecture, tackled in Section~\ref{sec:fullmimophy}, is designed using FPGA hardware module and LabVIEW FPGA software tool from National Instruments. In the next section, we illustrate the entire prototype setup of compact full duplex MIMO radios.
To characterize the actual performance in a changing environment, we designed the proposed full duplex MIMO PHY architecture, tackled in Section~\ref{sec:fullmimophy}, using FPGA hardware module and LabVIEW FPGA software tool from National Instruments. In the next section, we illustrate the entire prototype setup of compact full duplex MIMO radios.

%%%%%%%%%%%%%%%%%%%%%%%%%%%%%%%%%
%%%%%%%%%%%%%%%%%%%%%%%%%%%%%%%%%
\section{Prototype Setup}
\label{sec:prototypesetup}

Our compact full duplex MIMO prototype, shown in Fig.~\ref{fig:Hardware}-(a), has a two-node pair with a $2 \times 2$ configuration. The full duplex MIMO node consists of two parts -- the dual-polarized full duplex MIMO front-end as the analog canceler and the digital baseband PHY based SDR platform.

The full duplex MIMO PHY is implemented on two main processing modules, controlled by LabVIEW, a Xilinx Kintex~7 K410T FPGA in NI PXIe-7975 FlexRIO FPGA module~\cite{ni7975} and NI PXIe-8135 real-time (RT) controller equipped with a 2.3 GHz quad-core Intel Core i7-3610QE processor~\cite{ni8135}. Kintex~7 K410T FPGA has 63,500 slices and 28,620 Kb of block random access memory (RAM)~\cite{kintex7}. We mainly use slices in FPGA to implement from a basic arithmetic logic unit (ALU) to each independent, dedicated processing unit of full duplex PHY. As temporary storage spaces, the block RAMs are used to guarantee harmonious dataflow and prevent loss of data samples from among FPGA targets. These resources play a key role in canceling out the appropriate rebuilt self-interference from the Rx streaming data on time. We deploy the Xilinx intellectual property (IP) core defined in~\cite{hpFPGA} for FFT / IFFT operations and the finite impulse response (FIR) filter designs of low-pass filter, correlation filter for peak detection, and interpolator for channel estimation.  The RT controller is connected with FPGA modules through a chassis. The NI PXIe-1082 chassis~\cite{ni1082} offers reference clock sharing and data aggregation, for real-time signal processing between FPGA modules and the RT controller. The FPGA module is connected to an NI 5791R baseband transceiver.
The NI 5791R baseband transceiver features dual 130 MS/s ADC with 14-bit accuracy, and dual 130 MS/s digital-to-analog converter (DAC) with 16-bit accuracy and provides approximately 86~dB of dynamic range and coverage over the entire 100~MHz ISM band~\cite{ni5791}. As shown in~(\ref{eq:minasic}), an important specification is 1-dB compression point, $P_{1{\rm{dB}},R}$, of LNA because it determines the requirements of the minimum analog cancellation level to avoid self-interference from entering the nonlinear region. The LNA of NI 5791 has a 1-dB compression point of -15 dBm~\cite{ni5791}. The maximum power level (23 dBm) of a cellular or D2D user could not be supported by NI 5791. We thus utilized an off-the-shelf, additional wideband amplifier (ZVA-183W+~\cite{poweramp}) for the measurement.

%%%%%%%%%%%%%%%%%%%%%%%%%%%%%%%%%
%%%%%%%%%%%%%%%%%%%%%%%%%%%%%%%%%
\section{Experimental Evaluation}
\label{sec:phyeval}
In this section, we provide experimental results of our compact full duplex MIMO prototype in an indoor testbed. From our experimental results, we characterize the performance of a full duplex MIMO link depending on the Tx power of the full duplex transceiver, and the link quality between nodes. First, we introduce our experiment scenario for the measurement campaign.

   \begin{figure*}[t]
    \centering
    \includegraphics[width = 7.2in]{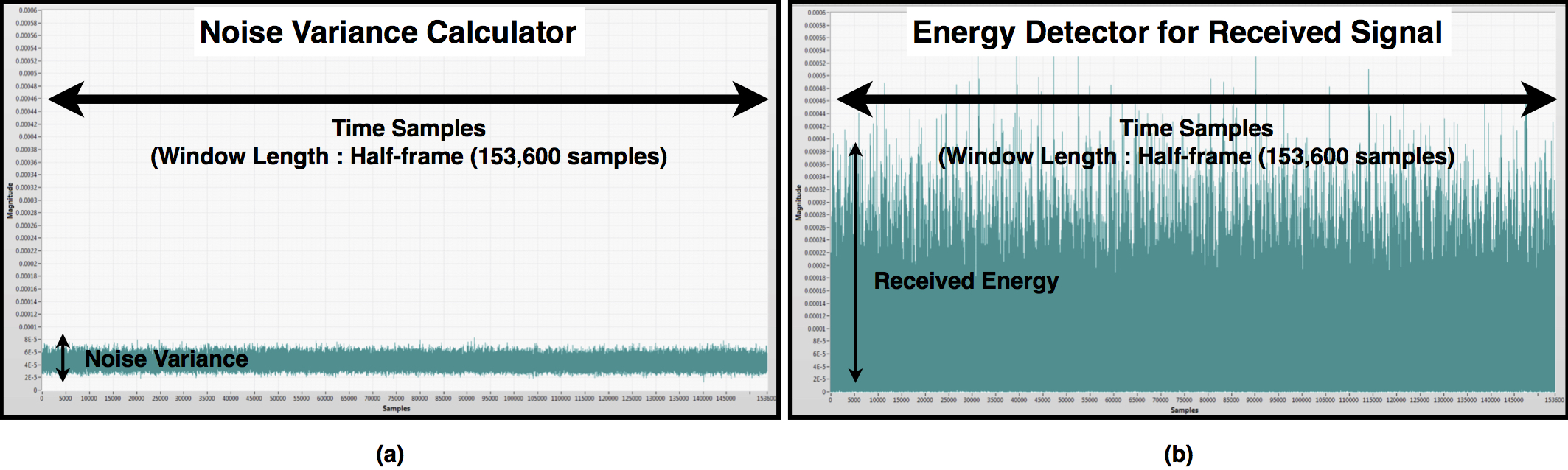}
    \caption{Snap shots of noise variance and received signal energy measurement in SDR platform for link quality mapping regarding indoor full duplex channel. }              
    \label{fig:Energy_Detec}
    \end{figure*}

%%%%%%%%%%%%%%%%%%%%%%%%%%%%%%%%%
%%%%%%%%%%%%%%%%%%%%%%%%%%%%%%%%%
\subsection{Experiment Scenario}
\label{subsec:expscen}
For the performance evaluation, the measurement campaign was conducted in various locations of Veritas Hall Building~C at Yonsei University, shown in Fig.~\ref{fig:Expscen}-(a). We wanted to measure and collect the most accurate performance results at as many spots as possible. Hence, we included the performance measurement unit that, in our prototype, calculated in real-time. Through this setting, with one of the two nodes fixed at some spot, we measured link qualities and performance values between the two nodes, using the other as a moving node. We were thus able to achieve the actual link quality and its performance value mapping for a real-world indoor network. Consider a signal model at time index $n$ to define link quality as follows. It is distinguished by the signal model defined in~(\ref{eq:signalmodel}) by index.
\begin{align}
\label{eq:signalmodelt}
{\textbf{y}}[n]=
{\textbf{H}}[n]{\textbf{x}}[n] +
{\textbf{z}}[n]
\end{align}
where ${\textbf{H}}[n] \in \mathbb{C}^{N{r} \times N_{t}}$ denotes the channel matrix between two nodes.  ${\textbf{x}}[n] \in \mathbb{C}^{N{t} \times 1}$  denotes Tx symbol vector and ${\textbf{z}}[n] \in \mathbb{C}^{N{r} \times 1}$ denotes noise vector at the Rx side.  Therefore, the link quality at $n$th time index is defined as,
\begin{align}
\label{eq:LQ11}
{{\rm{LQ}}_{n}}= \frac
{\mathbb{E} \left[
{
\begin{Vmatrix} 
{\textbf{H}}[n]{\textbf{x}}[n] 
\end{Vmatrix}_{F}^{2}
}
\right]}{\sigma_{\textbf{z}}^{2}[n]}= \frac
{\mathbb{E} \left[
{
\begin{Vmatrix} 
{\textbf{y}}[n]-{\textbf{z}}[n] 
\end{Vmatrix}_{F}^{2}
}
\right]}{\sigma_{\textbf{z}}^{2}[n]}
\end{align}
where $\sigma_{\textbf{z}}^{2}[n]$ represents the noise variance of $\textbf{z}[n]$. Assuming that the noise at the receiver is additive white Gaussian noise (AWGN), we have an approximated expression of link quality as,
\begin{align}
\label{eq:LQ11}
{{\rm{LQ}}_{n}}\simeq \frac
{\mathbb{E} \left[
{
\begin{Vmatrix} 
{\textbf{y}}[n]
\end{Vmatrix}_{F}^{2}
}
\right]-\sigma_{\textbf{z}}^{2}[n]}{\sigma_{\textbf{z}}^{2}[n]}
\end{align}
Before the measurement of link quality, we first calculated the average noise variance value during 20 half-frames (3,072,000 samples) using our testbed, with the transmitter off. Then applying that result, we performed the measurement campaign. In Fig.~\ref{fig:Energy_Detec}, we provide snap shots of the noise variance calculation and the energy detector of the received signal.

For the characterization of compact full duplex MIMO performance, we implemented an LTE-based conventional half duplex PHY and full duplex SISO PHY on the same SDR platform to utilize a performance benchmark. The measurement campaign regarding these testbeds was also performed in the same experiment scenario. As noted above, we followed the frame structure of the LTE downlink for half / full duplex transmission. Table~\ref{tb:Parameters} summarizes the parameters regarding the transmission of our compact full duplex $2 \times 2$ MIMO prototype in indoor experiments.

\begin{table}[ht]
\caption{Parameters for the full duplex $2 \times 2$ MIMO prototype.}
\label{tb:Parameters}
\centering
\begin{tabular} {c|c}
\hline 
\bfseries{Parameters} & \bfseries{Values}\\
\hline\hline 
Carrier Frequency & 2.52 GHz\\
Maximum Transmit Power & 23 dBm\\
Minimum Transmit Power & 0 dBm\\
Sampling Frequency & 30.72 MS/s\\
Subcarrier Spacing & 15 kHz\\
Bandwidth & 20 MHz\\
FFT Size & 2048\\
CP Length & 512 (Extended CP in LTE)\\
Number of Data Subcarrier & 1200 (Per OFDM symbol)\\
\hline
\end{tabular}
\end{table}

   \begin{figure}[h]
    \centering
    \includegraphics[width = 3.6in]{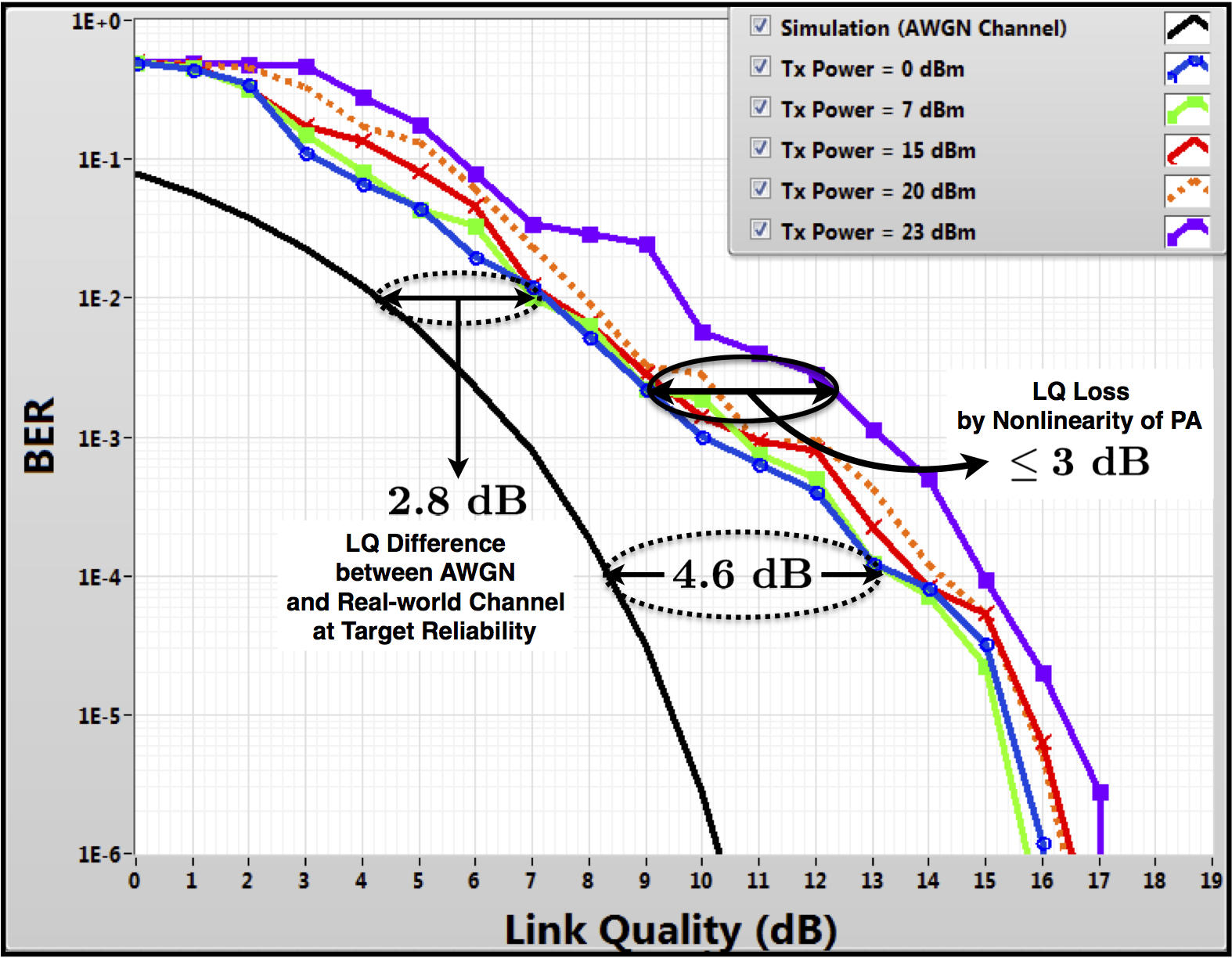}
    \caption{BER performance of Uncoded QPSK using our conventional LTE PHY prototype.}              
    \label{fig:HD_BER}
    \end{figure}
%%%%%%%%%%%%%%%%%%%%%%%%%%%%%%%%%
%%%%%%%%%%%%%%%%%%%%%%%%%%%%%%%%%
\subsection{Impact of PA Nonlinearity}
\label{subsec:impPAnon}
Let us first characterize the impact of PA nonlinearity depending on Tx power on the bit-error-rate (BER)  performance in our testbed. Because we use, as one of the main performance metrics, the BER for the characterization of compact full duplex MIMO radios in this paper, it would be meaningful to look into the impact of PA nonlinearity in terms of BER. To explore the nonlinearity impact of the only PA, for the experiment in this section we used the conventional LTE PHY prototype, operating in half duplex mode.

Fig.~\ref{fig:HD_BER} shows a comparison of the BER with different Tx power (dBm) values, $P_{T} \in \{0, 7, 15, 20, 23\}$ at the transmitter. 
As a comparison target, the simulation result under AWGN channel is attached. For simulation and experiments, the study used uncoded quadrature phase shift keying (QPSK). Link quality difference between AWGN channel, under the simulation, and real-world channel appears 2.8 dB, and 4.6 dB, respectively, at $\zeta_{t}(\%)=99.999$, and $99.9999$. $\zeta_{t}$ denotes the target reliability. In cases where $P_{T} \leq 15$, the impact of PA nonlinearity was not found over any of the link quality region. But in cases where $P_{T} \geq 20$, link quality degradation occurs over the entire link quality region. Also we observe that link quality loss is generated under 3 dB by PA nonlinearity in cases where the maximum Tx power of user $P_{T}=23$. In the following section, we explain these results regarding the impact of PA nonlinearity and the performance of a full duplex SISO / MIMO link.

    \begin{figure}[t]
    \centering
    \includegraphics[width = 3.6in]{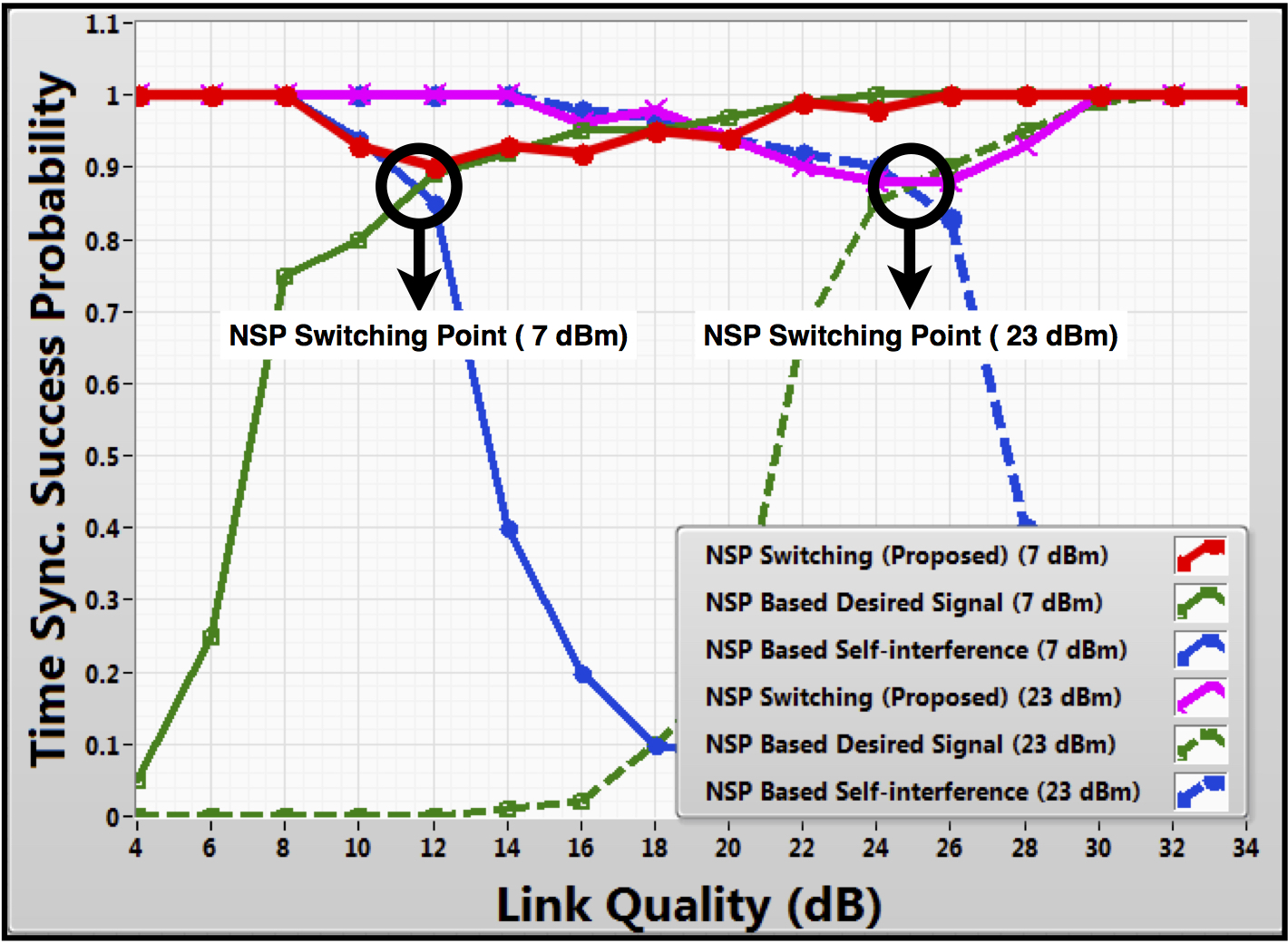}
    \caption{Success probability performance of full duplex MIMO timing synchronization according to different NSP index selection methods and Tx power values, i.e., $P_{T} \in \{7, 23\}$ of full duplex MIMO radio. The synchronization success probability by the proposed NSP switching method is achieved over 90 \% in case where $P_{T}= 7$ (red), $23$ (pink).}              
    \label{fig:SyncProb}
    \end{figure}

%%%%%%%%%%%%%%%%%%%%%%%%%%%%%%%%%
%%%%%%%%%%%%%%%%%%%%%%%%%%%%%%%%%
\subsection{Synchronization Success Probability}
\label{subsec:syncSP}
To initiate full duplex data exchange between the scheduled nodes, let us put the timing synchronization algorithm through its paces. Based on our proposed timing synchronization algorithm in Section~\ref{subsec:syncdesign}, we include a part of the compact full duplex MIMO prototype. As a result, we could measure the timing synchronization success probability as link quality in a planned indoor environment.

Figure~\ref{fig:SyncProb} shows the success probability performance of full duplex MIMO time synchronization according to different NSP index selection strategies. For the performance measurement, we consider two Tx power values (dBm), $P_{T} \in\{7, 23\}$. As noted above, when the link quality is low, there is higher synchronization success probability exploiting the PSS sent from own transceiver. The other way, when the link quality is in good condition, the PSS of the desired signal is more helpful. As shown in Fig.~\ref{fig:SyncProb}, we observe this trend in the case where $P_{T}=7$. Our testbed uses the crossing point for NSP switching when the link quality is approximately 12 dB. As a result, the proposed method enables a synchronization success probability in the case of Tx power being 7 dBm of over 90 \% in the link quality range between 4 and 34 dB. Also when the Tx power is 23 dBm, we achieve fine synchronization success probability performance. But one remarkable point is the place of the crossing point for NSP switching; this is located within a comparatively high link quality region. This means that, despite two full duplex nodes being in close proximity, they should still select the NSP index by the PSS of each self-interference in a direction opposite our regime. This occurs because of residual self-interference. Since the Tx power of a full duplex node is too high, in the case where its power exceeds the self-interference cancellation capability of compact full duplex MIMO radio, the crossing point appears in a higher link quality region by the amount of the residual interference.

    \begin{figure}[t]
    \centering
    \includegraphics[width = 3.6in]{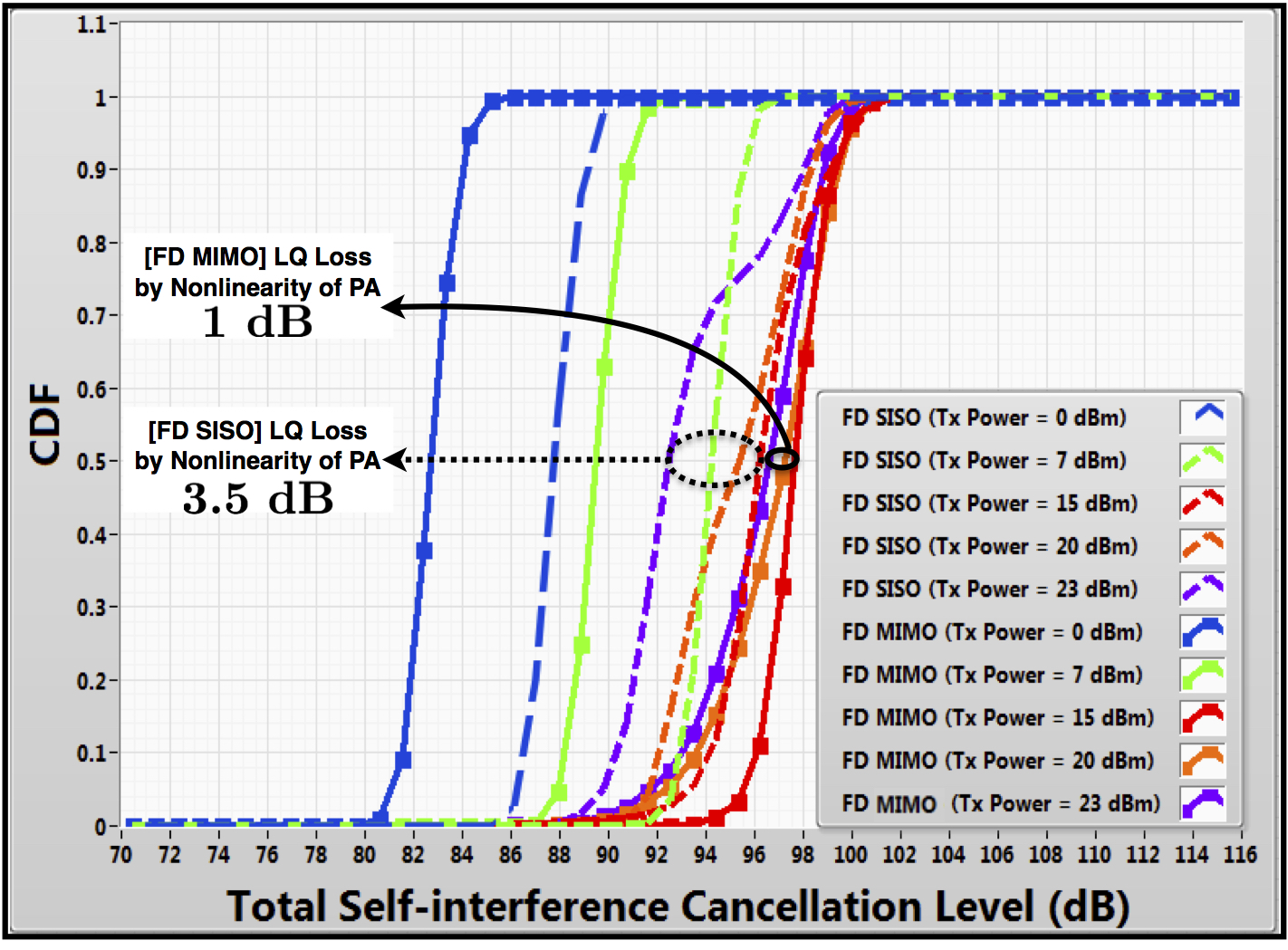}
    \caption{Cumulative density functions of total self-interference cancellation level according to different Tx power values, i.e., $P_{T}~\in~\{0, 7, 15, 20, 23\}$ of full duplex SISO (dot lines) / MIMO (solid lines) radios. The link quality loss by PA nonlinearity from Tx power level of 15 dBm to 23 dBm, full duplex SISO radio (3.5 dB) is larger than MIMO configuration (1 dB). }              
    \label{fig:SIC_CDF}
    \end{figure}

%%%%%%%%%%%%%%%%%%%%%%%%%%%%%%%%%
%%%%%%%%%%%%%%%%%%%%%%%%%%%%%%%%% 
\subsection{Self-Interference Cancellation Level}
\label{subsec:siclevel}
Let us provide the total self-interference cancellation capability of our compact full duplex MIMO prototype. The total self-interference cancellation level is defined as the addition of an analog cancellation level,~$\alpha$, and a digital cancellation level,~$\delta$, achieved at the maximum level of about 101 dB. For the characterization, we measure the total self-interference cancellation level according to the difference of Tx power in various indoor locations. For candidates of Tx power, we consider ${P_{T} \in \{0, 7, 15, 20, 23\}}$. For comparison, we also explore the total self-interference cancellation level of full duplex SISO radio, implemented with the same SDR platform, with the same dual-polarization-based full duplex front-end.

    \begin{figure}[t]
    \centering
    \includegraphics[width = 3.5in]{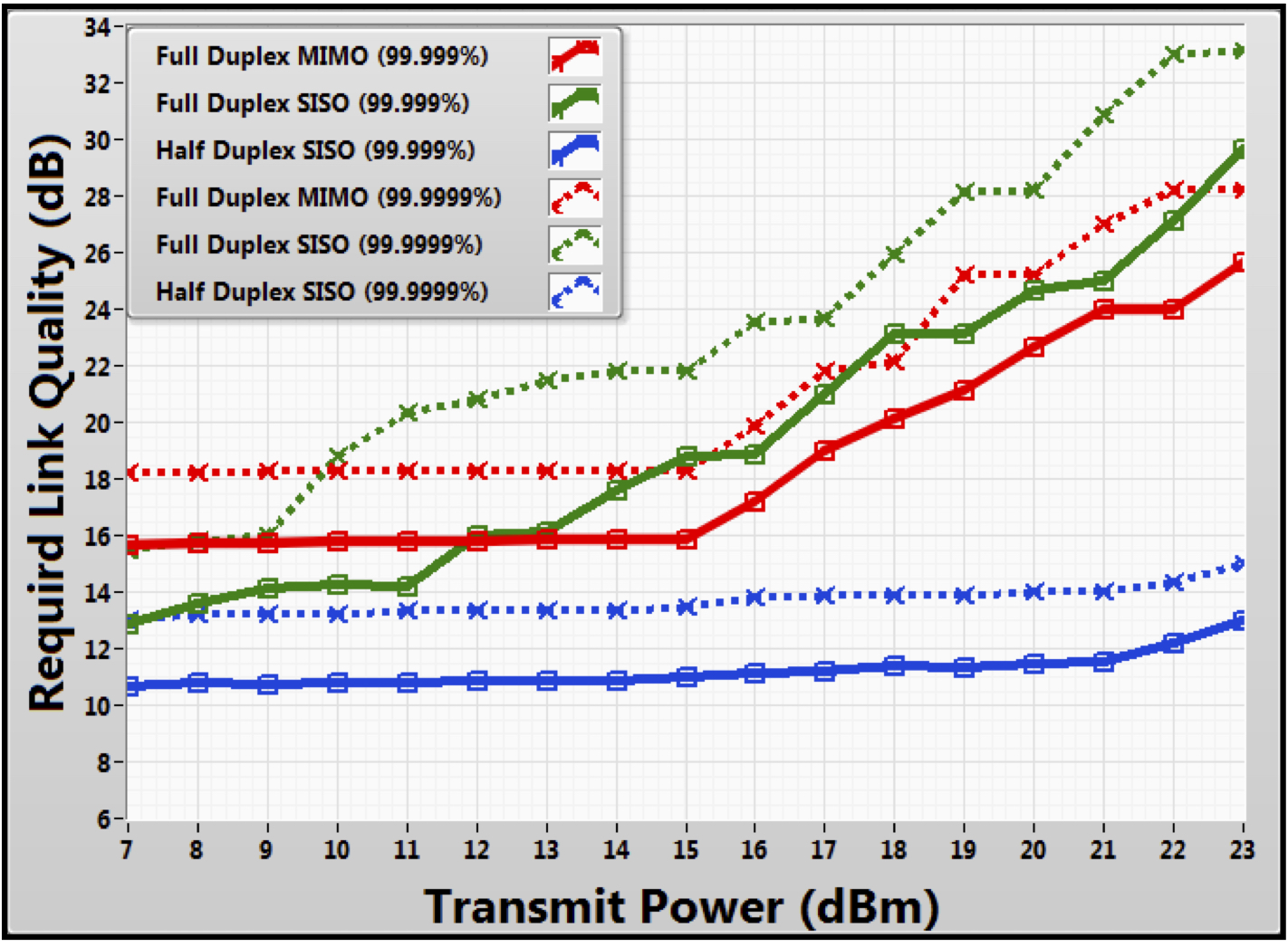}
    \caption{Required link quality according to Tx power of half / full duplex SISO / MIMO radios with a target reliability of $\zeta_{t}=99.999$\% (solid lines), and 99.9999 \% (dot lines) at each node.}              
    \label{fig:reqLQ}
    \end{figure}

Fig.~\ref{fig:SIC_CDF} shows a characterization of total self-interference cancellation levels of compact full duplex SISO / MIMO radio regarding different Tx power levels of full duplex radio. First, we observe a cumulative density function (CDF) value of 0.5. Since total self-interference cancellation levels are lower when Tx power levels are between 0 and 7~dBm, full duplex radio already cancels out self-interference to at the noise floor level, despite having the capability of canceling more self-interference. From 15 dBm to 23 dBm of Tx power, we observe that total self-interference cancellation levels of full duplex SISO / MIMO radios gradually drop by impact of PA nonlinearity. But we find that at a point value of 0.5, link quality loss of full duplex SISO radio is larger than that of full duplex MIMO radio. This may be interpreted as meaning that full duplex MIMO radios are less vulnerable to PA nonlinearity than is a SISO configuration at the maximum power level of 23 dBm. For example, on a full duplex $ 2 \times 2$ MIMO radio, Tx power is divided between two antennas for transmission using a power level of 23 dBm. A full duplex SISO radio, in the other hand, uses just the one antenna for all amounts of Tx power. Therefore, when each radio uses the same amount of Tx power, there is less inflowing power per PA of antenna.

%%%%%%%%%%%%%%%%%%%%%%%%%%%%%%%%%
%%%%%%%%%%%%%%%%%%%%%%%%%%%%%%%%%
\subsection{Required Link Quality}
\label{subsec:ReqLQ}

Let us consider the performance of compact full duplex MIMO from a reliability standpoint. The authors in~\cite{Osseiran2014} explained what is driving 5G research, describing scenarios for mobile and wireless communications.  In such scenarios, as one of the target requirements, the messages must be transferred with about 8 ms end-to-end delay on the application layer, with 99.999 \% reliability. 

In Fig.~\ref{fig:reqLQ}, we provide experimentally the required link quality according to Tx power when compact full duplex radios operate for latency improvement with a target reliability. We consider two kinds of target reliability values, i.e., $\zeta_{t} \in \{99.999, 99.9999\}$. For performance comparison, we also consider the results of full duplex SISO and half duplex SISO. In the case where a full duplex MIMO radio uses the maximum Tx power (23 dBm), the link quality must be approximately 26 dB and 28 dB, to achieve target reliabilities of 99.999 and 99.9999\% in an indoor environment.  The reason that the required link quality of full duplex MIMO is maintained when Tx power is under 15 dBm, is due to the self-interference of own radio to be cancelled. We observe that in the Tx power region the required link quality appears to be a rising curve. This increase of Tx power produces an excess of self-interference cancellation capability. One interesting point is that in the case of full duplex SISO the starting point of the rising curve is located at a lower Tx power level than in the MIMO case, and the slope of required link quality is steeper than in the MIMO case. As a result, the curves of the required link quality produce a crossing point at some Tx power value between full duplex SISO and MIMO. Such a finding may be interpreted as resulting from the difference in self-interference capability between full duplex SISO and MIMO. After all, the analog self-interference cancellation level of full duplex MIMO is higher than that of SISO. Also, the gentler slope of full duplex MIMO to that of full duplex SISO is due to the smaller impact of PA nonlinearity by the power separation between Tx antennas.

%Scheduling
%When we traced the link quality map of indoor environment, the moving node carried out the only transmission by the fixed power of 7 dBm.
%and the fixed node operated on full duplex mode. We measured 

\subsection{Throughput Performance}
\label{subsec:ReqLQ}
 \begin{figure}[t]
    \centering
    \includegraphics[width = 3.5in]{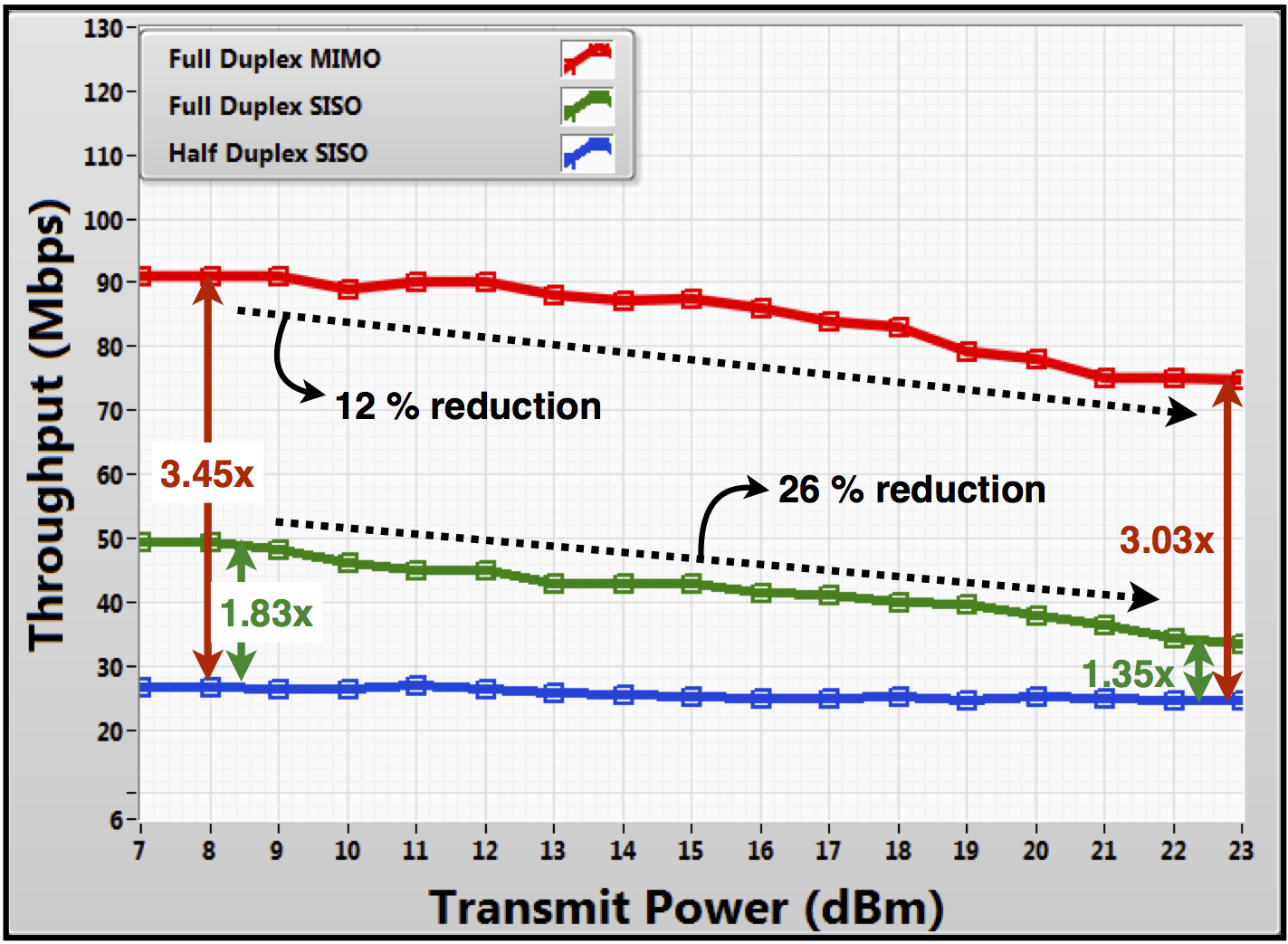}
    \caption{Achieved link throughput performance according to Tx power in an indoor mobile network with the transmission bandwidth of 20 MHz using QPSK as a fixed modulation scheme at each node.}              
    \label{fig:THPperformance}
    \end{figure}

As the last step in this section, so as to validate the application possibility of compact full duplex MIMO radios in D2D networks, let us investigate the link throughput performance in an indoor mobile environment. As noted in~Section~\ref{subsec:expscen}, one node was fixed at some location while the other node was used as a moving user to replicate a D2D scenario.

Fig.~\ref{fig:THPperformance} shows the experimentally achieved link throughput performance of compact full duplex $2 \times 2$ MIMO design compared to the full duplex SISO and conventional LTE SISO, according to Tx power. We placed the fixed node at 10 different locations-- represented by the red dots in Fig.~\ref{fig:Expscen}-- collected the bitrate which was provided as real-time results according to the location of the moving node. The link quality between the two nodes, ranged from 0 dB to 40 dB. Note that all of these throughput values account for the overhead in each frame structure. As shown in Fig.~\ref{fig:THPperformance}, our compact full duplex MIMO radio achieves an average throughput gain of approximately 3.45$\times$ over the conventional half duplex LTE SISO mode in an indoor mobile environment when Tx power is under 15 dBm. As the Tx power of a compact full duplex MIMO radio increases, the residual self-interference along with the PA nonlinearity incur an average throughput gain reduction of 12 \%. In spite of using the maximum power of a full duplex user, however, it achieves 3.03$\times$ more than the half duplex SISO mode. The average throughput gain reduction of full duplex SISO is higher than that of full duplex MIMO. This is because of lower self-interference cancellation capability and the higher impact of PA nonlinearity compared to compact full duplex MIMO, as discussed above.

%For application possibility in network scenarios where we mentioned, we measure the cancellation performance in various locations 

\section{Conclusion}
\label{sec:conclusion}
We have considered, from system design to their implementation, compact full duplex MIMO radios for users in D2D underlying cellular  networks. For the full duplex MIMO operation of a user, we have proposed a coupled structure, consisting of the following two components: a compact and power-efficient, dual-polarization-based analog solution and an LTE-based per-subcarrier digital self-interference canceler. The latter is robust to a frequency selective channel in a mobile network. We have also proposed an NSP index-switching algorithm to initiate full duplex communications via a timing synchronization considering the desired signal and self-interference of each user. By implementing this algorithm on our full duplex MIMO link prototype, we verified that the NSP index switching enabled the achievement of high timing synchronization success probability in a full duplex link. A major implication from this experimental evaluation is that, even when a user utilizes the maximum Tx power, full duplex MIMO radios designed for users are still able to acquire throughput gain in an indoor mobile network. Furthermore, if a power control strategy is applied according to link quality between full duplex nodes, we showed that throughput improvement is also achieved by full duplex operation, with target reliability for each node.

We believe this paper provides the possibility of being able to use full duplex communications at not only the base station but also the user node. This compact full duplex MIMO architecture can be deployed for higher throughput and lower latency performance with many wireless applications. Its standout features  include heterogeneous network coexistence, spectrum sharing, and power control.

\ifCLASSOPTIONcaptionsoff
  \newpage
\fi

\renewcommand{\baselinestretch}{1.0}
\bibliographystyle{IEEEtran}
\bibliography{reference_IEEE2016} % file name

\clearpage

\end{document}